\documentclass[preprint,11pnt]{aastex}
\usepackage{epstopdf}
\usepackage{multirow}
\usepackage{arydshln}
\RequirePackage{mathrsfs}

\newcommand{\oxygen}{O$_2$}
\newcommand{\water}{H$_2$O}

\slugcomment{Accepted to PASP}

\shorttitle{Sub-Millimagnitude Photometry from the Ground}
\shortauthors{Mann et al.}

\begin{document}

\title{Ground-Based Submillimagnitude CCD Photometry of Bright Stars Using Snapshot Observations}

\author{Andrew W. Mann\altaffilmark{1}, 
Eric Gaidos\altaffilmark{2}, 
Greg Aldering\altaffilmark{3}}
\email{amann@ifa.hawaii.edu}

\altaffiltext{1}{Institute for Astronomy, University
  of Hawaii, Honolulu, HI 96822} 
\altaffiltext{2}{Department of Geology and Geophysics,
  University of Hawaii, Honolulu, HI 96822}
\altaffiltext{3}{Lawrence Berkeley National Lab, CA 94720}

\begin{abstract}
We demonstrate ground-based submillimagnitude ($< 10^{-3}$) photometry of widely separated bright stars using snapshot CCD imaging.  We routinely achieved this photometric precision by (1) choosing nearby comparison stars of a similar magnitude and spectral type, (2) defocusing the telescope to allow high signal ($> 10^7$~e$^{-}$) to be acquired in a single integration, (3) pointing the telescope so that all stellar images fall on the same detector pixels, and (4) using a region of the CCD detector that is free of nonlinear or aberrant pixels.  We describe semiautomated observations with the Supernova Integrated Field Spectrograph (SNIFS) on the University of Hawaii 2.2m telescope on Mauna Kea, with which we achieved photometric precision as good as $5.2\times10^{-4}$ (0.56~mmag) with a 5 minute cadence over a 2~hr interval.  In one experiment, we monitored eight stars, each separated by several degrees, and achieved submillimagnitude precision with a cadence (per star) of $\sim$17 minutes.  Our snapshot technique is suitable for automated searches for planetary transits among multiple bright stars.
\end{abstract}


\section{Introduction}\label{sec:intro}

High-precision stellar photometry is used to detect planets orbiting other stars, exploiting the phenomena of a transit \citep{2000ApJ...529L..45C}, a secondary eclipse \citep{2007prpl.conf..701C}, or a microlensing event \citep{2004ApJ...606L.155B}.  Although photometry from space can be vastly more precise - {\it Kepler} achieves $\sim2\times10^{-5}$ in 30 minutes with $V \sim 10$ stars \citep{2010ApJ...713L..79K} - the preponderance of astronomical resources are on the ground and must contend with absorption, scattering, and scintillation by the atmosphere.  Ground-based searches have been able to detect Neptune- to Jupiter-sized planets on close-in, transiting orbits around solar-mass stars \citep[e.g.,][]{2000ApJ...529L..41H,2006PASP..118.1407P, 2007A&A...472L..13G}, a multi-Earth-mass planet around an M dwarf \citep{2009Natur.462..891C}, and planets on more distant orbits around M stars: \citep[e.g.,][]{2010ApJ...720.1073G}.

\begin{figure}
\centering
\includegraphics[width=\textwidth]{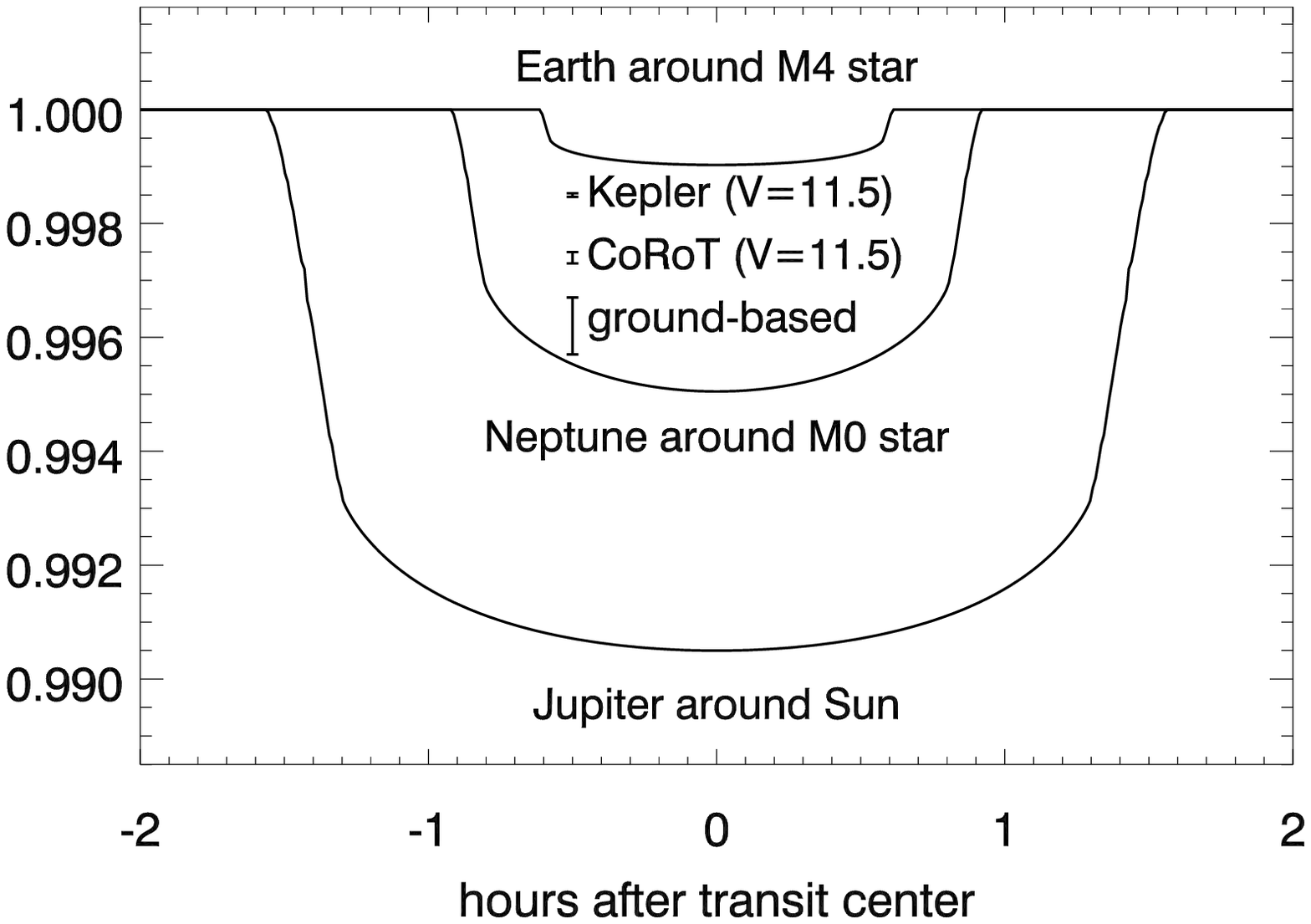}
\caption{Theoretical normalized light curves of stars transited by planets on 4 day orbits, compared with the photometric precision obtained from space - {\it Kepler} 30 minute \citep{2010ApJ...713L.120J} and COROT \citep{2009A&A...506..425A} - and a practical limit from the ground ($5\times10^{-4}$). Limb darkening is included as described in \citet{2000A&A...363.1081C}.  A stellar mass-radius relation $R_* \sim$M$_*^{1.06}$ (appropriate for M dwarfs) is used.  Ground-based photometry can be sufficiently stable to detect and characterize super-Earth- to Neptune sizes planets around M dwarf stars. \label{fig:lightcurve}}
\end{figure}

Relative photometric stability of a few millimagnitudes on the timescale of minutes to hours is sufficient to detect the transit of a Jupiter-sized planet (Fig.~\ref{fig:lightcurve}).  However, superior stability is required to resolve degeneracies between the unknown impact parameter, size of the star, and limb-darkening effects, and to search for transit timing variation indicative of other, unseen planets \citep{2009AJ....137.3826W,2010MNRAS.408.1689S}.  Furthermore, high-precision Doppler studies are now detecting Earth- to Neptune-mass planets on short-period orbits \citep{2009A&A...507..487M,2009ApJ...696...75H,2010A&A...512A..48L,2010ApJ...708.1366V}.  The transit depth of these objects, even around smaller M stars, is at most a few millimagnitudes (Fig.~\ref{fig:lightcurve}).  Signal-to-noise ratio (S/N) can be improved by phasing data with the known or hypothetical orbital period \citep{2006ApJ...652.1715H, 2007AJ....133...11W, 2007A&A...472L..13G}, but the gains from phasing (and binning) are limited by time-correlated (red) noise, which has a $(1/f)^{\gamma}$ varying power spectral density, and does not decrease as the square root of the number of observations \citep{2006MNRAS.373..231P}.  Rather, photometric stability better than $10^{-3}$ ($\sim$1 mmag) in a single observation is required.  

To surpass a precision of $10^{-3}$, $\sim$10$^7~e^{-1}$  must be acquired per integration.  However, the dynamic range of optical CCD detectors limit the number of electrons in a single pixel to $\sim$10$^5$.  One method to avoid exceeding the linear range is to read out the detector at a rate of many hertz and to co-add the signal from multiple reads with low read noise \citep{2006PASP..118.1550S}.  Another is to spread the signal over hundreds of pixels by shaping the point-spread function (PSF) with orthogonal charge transfer \citep{2003PASP..115.1340H, 2005PASP..117.1187H, 2005PASP..117..281T,   2009ApJ...692L.100J} or by defocusing the telescope \citep[e.g.,][]{2009A&A...496..259G,2009AJ....137.3826W,2009MNRAS.396.1023S}.  Longer integration times reduce scintillation noise and lower the fraction of time spent reading out the detector.

The deleterious effects of the atmosphere can be reduced by performing differential photometry with one or more comparison stars observed in the same field of view as the target star.  That correction's accuracy will be limited by the signal from the comparison star (or collection of comparison stars).  This technique commonly achieves photometric stability better than 1~mmag on ground-based observations of variable stars and exoplanet transits \citep{2004A&A...424L..31M,2005PASP..117..281T,2005AJ....130.2241H, 2009ApJ...692L.100J,2009A&A...496..259G,2009AJ....137.3826W, 2009MNRAS.396.1023S,2010MNRAS.408.1680S}.

A larger telescope can collect more signal but, all else being equal, its detector will have a smaller field of view with which to include a suitable comparison star.  For a given $f$-number, the field of view of a telescope with a specified imaging detector is inversely proportional to the aperture $D$.  The number-flux relation for nearby stars has an approximate slope of $-1.5$, thus for a given integration time, the signal from the brightest comparison star in the field of any target star scales as $D^2\times D^{-4/3}$, or $D^{2/3}$ (not considering binaries).  In principle, there is a modest advantage with larger telescope aperture.  However, for a typical CCD field of view of 10\arcmin, the probability of an equally bright comparison star appearing in the same field becomes large (i.e., 80\% at a galactic latitude of 30$^{\circ}$) only by an apparent magnitude of 13 \citep{2007hsaa.book.....Z}.  This brightness is at the limit of current Doppler techniques, and would exclude those systems most amenable to follow-up observations.

If a more suitable comparison star can be found in a separate, but nearby field, observations of the target and reference can be obtained by slewing the telescope between the two positions, and the exposure time can be adjusted to acquire equal signal from each star.  The comparison star could also be the target of a transit search.  These sequential observations will be neither simultaneous nor through the same column of the atmosphere.  This technique has been employed for decades with single-channel photomultiplier tubes (PMTs) and has been applied to studies of (and searches for) exoplanet transits and stellar variability, commonly taking advantage of automatic telescopes \citep[e.g.,][]{1987PASP...99..660G, 1991PASP..103..221Y, 1999PASP..111..845H, 2000ApJS..130..201H, 2008PASP..120..992G}.  One of the seminal articles on the subject, \citet{1991PASP..103..221Y}, describes the limits of photometry when slewing between stars and making use of high-speed robotic telescopes.  The authors suggest that the limit of ground-based automated photometry is $\sim$1 mmag when all sources of error are properly handled.  \citet{1999PASP..111..845H} and \citet{2000ApJS..130..201H} use automatic photoelectric telescopes, all less than 1~m in aperture, and achieve photometric precision $\gtrsim$1~mmag for single observations by observing a set of approximately four stars at a time, of which three are comparison stars.  They claim their precision is limited more by variability of comparison stars than by noise inherent to the technique.  So far, none of these efforts have been able to achieve consistent precision of less than 1 mmag.

By combining the single-channel PMT techniques with a high-performance CCD, we have routinely achieved submillimagnitude relative photometry with respect to one or more comparison stars under photometric conditions.  Unlike with PMTs, we are able to do simultaneous sky measurement and can take advantage of high quantum efficiencies.  We slew between multiple stars of similar spectral types and apparent magnitude, but separated by several degrees on the sky.  Our snapshot method combines (1) semiautomated observations while rapidly slewing between target and comparison stars, (2) telescope defocusing to achieve the requisite photon counts, and (3) precision CCD photometry practices (discussed later).  Here, we present data obtained using the SuperNova Integral Field Spectrograph (SNIFS) on the University of Hawaii UH2.2~m telescope at Mauna Kea Observatory.  In \S~\ref{sec:obs} we describe our observations, including the instrument, telescope, and targets.  In \S~\ref{sec:results} we quantify the major sources of error, and in \S~\ref{sec:minimization} we discuss our efforts to mitigate these sources of noise.   In \S~\ref{sec:noisemodel} we combine our noise sources using a simple noise model and compare with the noise from our observations.  In \S~\ref{sec:discussion} we discuss our findings in the form of guidelines for other observers, and we conclude with a simulated example of the technique's intended application; a search for shallow ($\gtrsim$2~mmag) planetary transits around stars that exhibit a Doppler signal.

\section{Observations} \label{sec:obs}

SNIFS is mounted on the south bent Cassegrain port of the UH 2.2~m telescope located atop Mauna Kea.  SNIFS is an integral field spectrograph \citep{2002SPIE.4836...61A,2004SPIE.5249..146L} originally designed for spectrophotometric observations of Type Ia supernovae for the Nearby Supernovae Factory project \citep{2010AIPC.1241..259P}. SNIFS consists of blue and red spectrograph channels, along with an imaging channel, mounted behind a common shutter. The imaging channel consists of two E2V $2048\times4096$ CCD detectors (CCD44-82-B23), one a high-quality science-grade device dedicated to imaging and the other an excellent engineering-grade device used for guiding.  Each 15 $\mu$m pixel subtends $0.137\arcsec$ and the field of view is 9.35\arcmin on a side.  The gain for the relevant amplifier of the imaging detector is 1.58~e$^{-}$/DN and the readout noise is 4~e$^{-}$, or 6~e$^{-}$ if electronic pickup noise is included.  The telescope control system is accessed by the SNIFS instrument computer, allowing the observer to run scripts that include telescope slews and offsets for target acquisition, focus changes, and guiding as described in \citet{2008SPIE.7016E..50A}. 

Our target and comparison stars are bright ($V <$ 13) late K and early M main-sequence stars that are the targets of a combined Doppler-transit planet search \citep{2011AJ....142..138L}. These stars are drawn from the proper-motion-selected SUPERBLINK catalog survey based on $V$-$J$ color and color-magnitude relationships \citep{2005AJ....130.1680L}.  Candidate targets are first screened using low-resolution spectra to confirm spectral type, determine metallicity, and remove those with strong H$\alpha$~emission \citep[potentially active or flaring stars,][]{2009AJ....138..633K}.  Stars are then monitored for barycenter motion with the High Resolution Spectrometer on the Keck I telescope \citep{2010PASP..122..156A}.  Targets showing significant radial velocity variation are then monitored for transits.  The full catalog comprises more than 13,000 M and K dwarf stars, and thus the nearest available comparison star from the catalog is usually no more than 2--3$^{\circ}$ away.

Observations were obtained between 2010 June 22 and 2011 January 15, usually within 7 days of a full Moon.  Photometric observations presented here usually took up approximately half of each observing night.  Observing conditions are presented in Table~\ref{tab:table1}.  Seeing was measured each time the telescope was focused: it is the full width at half-maximum of the PSF at optimal focus, interpolated between discrete focus settings.  Data on atmospheric extinction were obtained from the SkyProbe of the Canada-France-Hawaii Telescope (CFHT), which images Tycho stars around the boresight of that telescope through B and V filters, and determines extinction by comparing the observed signals to reference values \citep{2009PASP..121..295S}.  We report the median, standard deviation, and 95 percentile values of all the measurements in a night, excluding the first 10 and last 10 observations made near sunset and sunrise.  The SkyProbe data often contain high-extinction artifacts created when the telescope slews, so we check the SkyProbe data against videos from the UH~2.2m all-sky camera\footnote{http://uh22data2.ifa.hawaii.edu/public/allsky/index.php}, which can provide visual confirmation of clouds.  While extinction during the nights of June 22-24, July 27-29, and January 15 was low and constant (extinction of $\lesssim$0.2 mag is considered photometric), extinction during June 27-30 and January 15 was more variable; variation in extinction is more detrimental to relative photometry than the absolute value of the extinction.  Only June 30 showed evidence of (thin) clouds in the all-sky camera, predominately during the first few hours of the night.  We also report the estimated precipitable water vapor (PWV) above Mauna Kea Observatories for each night, based on the mean of the 225 GHz optical depth ($\tau_{225}$) measurements from the Caltech Submillimeter Observatory, and using the conversion PWV$ = 20\times(\tau_{225}-0.016)$mm \citep{1997Icar..130..387D}.  These values are germane because of H$_2$O absorption features between 6000 and 10,000~\AA, i.e., in the SDSS~$r$, $i$, and $z$ passbands. .

A single observation cycle consisted of (1) the acquisition of a smaller (800 $\times$ 800) image to locate the star after a slew, (2) calculation of the centroid of the focused stellar PSF, and the required offset to place the star at the desired location on the CCD, while at the same time (3) defocusing of the telescope by a specified amount, (4) acquisition of the image for photometry, (5) slewing to the next target while the image is read out, and (6) refocusing the telescope.  To minimize readout time, only the bottom 1024 rows of each CCD/amplifier combination were used to image the target.  For short slews, a single observation cycle requires $\sim$2~minutes to complete.  Nearly all observations were performed through the SDSS~$z$ passband, although some were performed with SDSS~$r$ \citep{1996AJ....111.1748F}.

\begin{figure}
\centering
  \includegraphics[width=\textwidth]{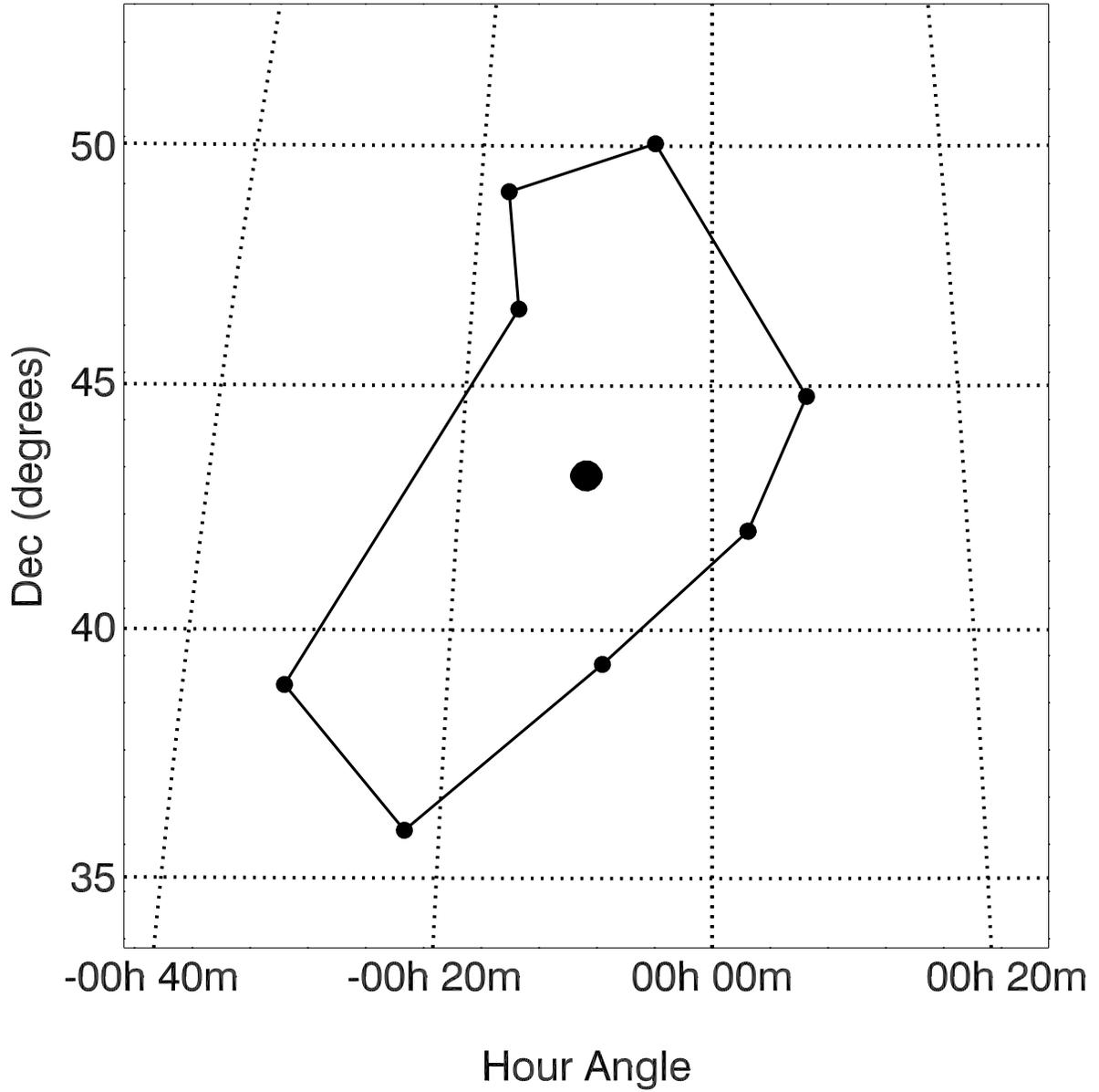}
  \caption{Aitoff projection of trajectory of telescope pointings for constellation observations, designed to minimize slew time.  The circle is approximately the size of the full Moon.  Although a given target is usually several degrees away from the next target, the slew time is not significantly longer than the readout time ($\sim$20~s).  The constellation technique enabled us to concurrently search for transits around eight target stars. \label{fig:skymap}}
\end{figure}

Observations are done in one of four different semiautomated modes, (1) a target with a comparison star(s) in the same field of view, (2) single-set snapshot, consisting of a single set  of stars (usually one target and one to two comparison stars) of similar spectral type and magnitude, and permitting the shortest cadence, (3) survey snapshot, where we observe two or more sets, with the stars in a set separated by less than a few degrees, but with the sets separated by as much as $30^{\circ}$, and (4) constellation mode, where we repeat a sequence of observations of a large number of neighboring stars in a circuit, each star serves as a potential comparison star for every other star in the constellation (Fig.~\ref{fig:skymap}).  

\section{Sources of Noise}\label{sec:results}
When the target and comparison stars are in the same field of view, differential photometry between the two is performed through nearly the same air column.  In snapshot photometry of stars separated by a degree or more, observations are neither simultaneous nor made through the same path in the atmosphere.  Inhomogeneities and atmospheric transmission fluctuations (ATFs) will produce errors, but these can be mitigated by judicious choices of passband, allowable air mass, and comparison star, as well as information about the atmosphere itself \citep{2007PASP..119.1163S}.  Here, we discuss each noise source, including an estimate of the amount of noise from each source in our observations.  We discuss some basic strategies for minimizing these sources of error in \S~\ref{sec:minimization}, and in \S~\ref{sec:noisemodel} we combine all noise sources to predict the noise level for each observation to be compared with the actual precision.

To compare with snapshot observations, we twice conducted an experiment in which two stars of similar spectral type and $z$-band magnitude that fortuitously fell within the same field of view were continuously observed (cadence, 2.4, 5.5 minutes).  The rms of the photometry was $9.7\times10^{-4}$ and $8.4\times10^{-4}$  (Table \ref{tab:table2}).  Noise from these observations is primarily due to scintillation, noise from the detector (see \S \S \ref{sec:scintillation} and \ref{sec:detectorbeh}, respectively), and Poisson noise.  In these experiments the comparison stars were of comparable fluxes as the target (total counts $\sim$1$\times10^{7}$~e$^{-}$).  Total signal detected in both tests was similar to that of our snapshot observations, where median signal is $\sim$1.4$\times10^{7}$~e$^{-}$ for both target and comparison stars, yielding Poisson noise of $\sim$2.7$\times10^{-4}$ for each integration.  The precision for these experiments was not better than that for our snapshot experiments.  

\subsection{Scintillation Noise}\label{sec:scintillation}
Scintillation noise from a telescope on Mauna Kea (4200~m elevation) is approximately
\begin{equation}\label{equation:scint}
  \sigma_s \approx 1.6z^{7/4}D^{-2/3}t^{-1/2}\left(\frac{\lambda}{5500}\right)^{-7/12} {\rm mmag},
\end{equation}
where $D$ is the diameter of the telescope in meters, $z$ is the air mass, $t$ is the integration time in seconds, and $\lambda$ is the wavelength in angstroms \citep{1993Obs...113...41Y, 2006obas.book.....B}.  Equation (\ref{equation:scint}) is only approximate, as scintillation also varies seasonally, with the speed and direction of the wind \citep{1974ApJ...189..587Y, 1998PASP..110.1118D, 2006obas.book.....B} and with conditions in the upper atmosphere \citep{1996PASP..108..385H}.   For example, the true value of the air mass exponent is $\sim$2 when observing in the same direction as the wind and is $\sim$1.5 when observing perpendicular to it,  and the wavelength dependency vanishes for larger telescopes \citep{1981PrOpt..19..281R}.  More rigorous methods of calculating scintillation can be found in \citet{2006PASP..118..924K} and \citet{2011AstL...37...40K}, but with proper modification of the air mass term for wind direction and no wavelength dependency, equation (\ref{equation:scint}) is consistent within $20\%$ of scintillation measurements from Mauna Kea \citep{1982ApOpt..21.1196D, 1993AJ....106.2441G, 2008JApMC..47.1140C} under conditions similar to those of our observations.  With wind modifications and without the wavelength term, equation (\ref{equation:scint}) gives a median value per star of $2.4\times10^{-4}$ for noise due to scintillation in our observations.  It becomes the dominant contribution to the error budget at exposure times if less than 15~sec.

\subsection{Extinction}\label{sec:extinction}
{\it First-order extinction:}
Our target and comparison stars, although close on the sky, will have a nonnegligible difference in air mass.  For an observation of stars lasting several hours, the differential air mass of the target and comparison star(s), $\chi = X_{targ} - X_{comp}$, will change.  For small values of $\chi$, and assuming no significant change in the extinction coefficient, the fractional change in the normalized relative flux will be $\delta f \approx 0.4\ {\rm ln}(10) E(\lambda)\Delta \chi$, where $E(\lambda)$ is the extinction coefficient in magnitudes and air mass and $\Delta \chi$ is the change in air mass difference over the course of an observation.   A typical value of $\Delta \chi$ in 2~hr for a well-selected comparison star is $0.003$.  Assuming an extinction coefficient for photometric conditions of $\sim$0.04~mag air mass$^{-1}$ in SDSS~$z$ \citep{1979PASP...91..571M, 2009MNRAS.394..675H}, the change in flux of the target star relative to the comparison star will be $\sim$10$^{-4}$.  Even for comparison stars within 5$^{\circ}$ of the target, values of greater than 0.01 for $\Delta \chi$ are possible, and the resulting change in flux could be as much as $\sim$4$\times10^{-4}$.  If large values of $\Delta \chi$ combine with higher extinction coefficients (i.e., nonphotometric or high mean PWV), the systematic effect can be larger than $\sim$10$^{-3}$.  However, the resulting trend is systematic and can be removed with $\sim$10$^{-4}$ precision provided that the extinction coefficient is measured (or known) to better than $0.01$~mag air mass$^{-1}$.  

{\it Second-order extinction:} If the target and comparison stars do not have the same spectral energy distributions over the passband $\Delta \lambda$, changes in the extinction by the atmosphere with air mass and time will produce trends in their relative signals. The effect will be proportional to $\Delta \lambda^2$.  Second-order extinction error was analyzed by (among others) \citet{1991PASP..103..221Y} and \citet{2001PASP..113.1428E}.  \citet{1991PASP..103..221Y} estimate the size of the effect by observing two stars differing in $B$-$V$ by 0.3 mag and find that the difference in $B$ extinction of the two stars is $\sim$7 mmag air mass$^{-1}$, although other estimates are notably smaller: e.g., \citet{2001PASP..113.1428E}.  The color effect is smaller in redder passbands, with stars of later spectral type, and with narrower bandwidth filters.  In the case in which the spectral energy distributions of the target and comparison stars are known, a correction can be made.  

\begin{figure}
\centering
  \includegraphics[width=\textwidth]{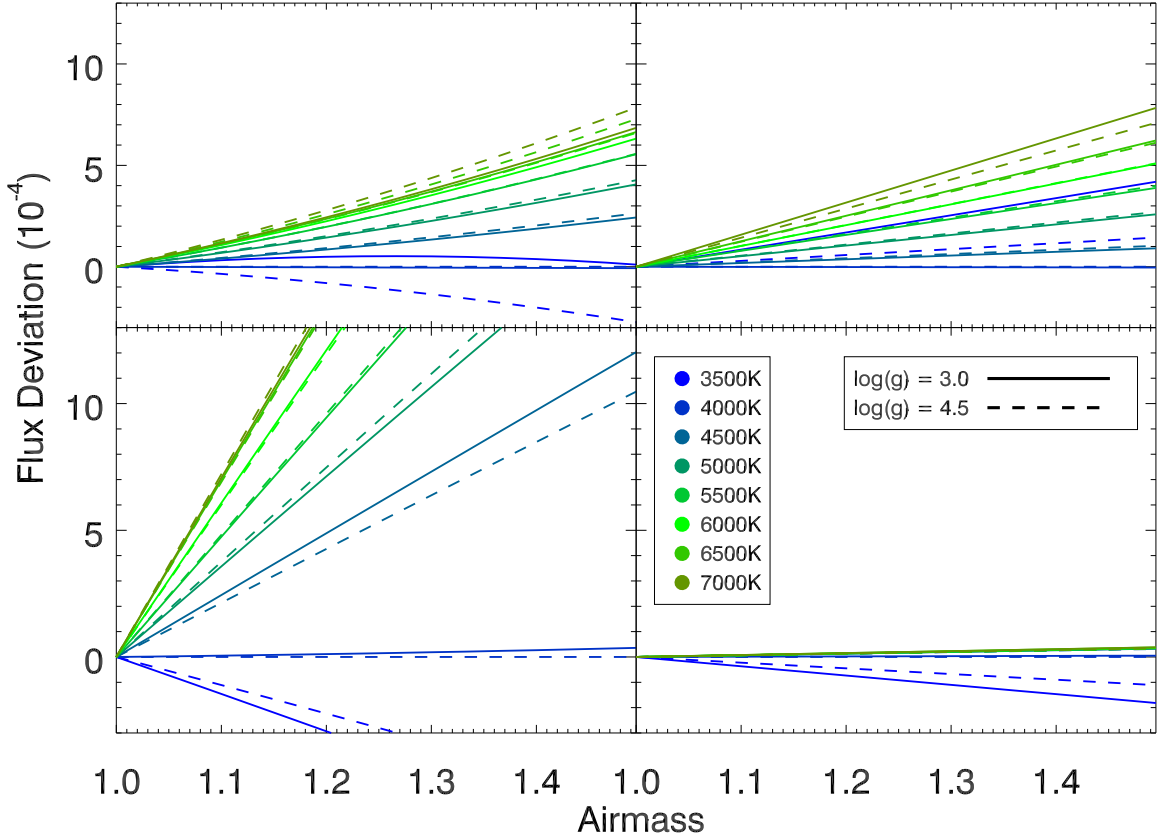}
  \caption{Photometric error in observations with the SDSS~$r$ (top left), SDSS~$z$ (top right), $B$ (bottom left), and narrow $z$ filters produced by a difference in spectral type between comparison and target star as a function of air mass (both using the SNIFS detector).  The target star has an effective temperature of 4000~K and $\log g = 4.5$. \label{fig:specerror}}
\end{figure}

We estimate second-order extinction errors with respect to a typical target star in our observations (4000~K and log~g = 4.5).  We convolve model spectra of \citet{1991sabc.conf..441K} with the transmission curve of the SDSS~$r$, SDSS~$z$ \citep{1996AJ....111.1748F}, $B$, or a narrow custom $z$ (J. Johnson 2011 private communication) filter, the quantum efficiency of the EV2 detector, and a model of atmospheric transmission.  Fig.~\ref{fig:atmosfilters} shows the profile of each of these filters multiplied by the SNIFS transmission and with an approximation of the atmosphere above Mauna Kea.  Our atmosphere model is based on low-resolution spectra of standard stars taken with SNIFS, combined with the HITRAN software \citep{2009JQSRT.110..533R} in regions where our spectra are incomplete or contaminated by features of our standards.  Fig.~\ref{fig:specerror} shows the expected change (systematic error) in flux ratio ($F_{target}/F_{comparison}$, normalized to 1), as a function of air mass, for the SDSS~$r$, SDSS~$z$, $B$, and narrow $z$ filters.  We assume that the target and comparison stars are observed through identical air mass.  We find the size of the effect in the $B$ band to be approximately 2.5 mmag air mass$^{-1}$ for a comparison star with (B-V) color difference of 0.3, which is significantly smaller than that found by \citet{1991PASP..103..221Y}.  The difference is likely due to the difference in atmosphere above the observatories (Mauna Kea versus Mount Hopkins), different instrument profiles, and the choice of later spectral type stars for our calculations.  For our observations in the SDSS~$z$ filter. we calculate a median value of only $4.5\times10^{-5}$, which is a benefit of choosing comparison stars of a similar spectral type to the target star and working at longer wavelengths, where the differences in stellar spectra are smaller.

\subsection{Short-Term Atmospheric Transparency Variations}\label{sec:transvar}
Snapshot observations are not simultaneous ($\gtrsim 2$ min lag), and variations in the atmosphere will produce error.  \citet{1994SoPh..152..351H} find the average power
spectrum of transparency fluctuations in the atmosphere above Mauna Kea to obey
\begin{equation}\label{eqn:atmospower}
\log P(\nu) = - 9.84 - 1.50 \log \nu,
\end{equation}
where $\nu$ is the frequency in hertz.  Equation (\ref{eqn:atmospower}) is based on an average over 691 clear days.  An individual night could have significantly higher or lower ATFs, which makes this source of noise particularly difficult to estimate for any given observation.  Regardless, the spectrum contains no coherence time within the timescale of our observations, and the photometric stability improves with decreasing time between target and comparison star observations (shorter cadence), although with diminishing returns.  We make use of equation (\ref{eqn:atmospower}) to estimate the noise from atmospheric variations between observations of target and comparison stars.  In general, the variance due to ATF is
\begin{equation}\label{eqn:atmosfluct1}
\sigma_{ATF}^2 = \left<\left|\int_0^{t'} \! D(t) S(t) dt\ \right|^2\right>,
\end{equation}
where $D(t)$ and $S(t)$ are functions that describe the transparency fluctuations and integration windows respectively, and $<>$ represents the time-averaged expectation.  
We assume that our observations behave as a function of the form
\begin{equation}
   S(t) = \left\{
     \begin{array}{llr}
       1 & : t \in T_{targ}\\
       -1 & : t \in T_{comp} & ,\\
       0 & : t \notin  T_{targ} \hspace{1 mm} \bigcup \hspace{1 mm} T_{comp}
     \end{array}
   \right.
\end{equation}
where $T_{targ}$ and $T_{comp}$ are the times over which we are integrating on the target and comparison star respectively.  We can represent $S(t)$ and $D(t)$ as Fourier transforms $\mathscr{D}(\omega)$ and $\mathscr{S}(\omega)$ and substitute into equation~\ref{eqn:atmosfluct1}.  After some simplification, we find

\begin{equation}\label{eqn:atmosfluct2}
\sigma_{ATF}^2 \approx \left<\int_0^{\omega'} \! (\mathscr{D}(\omega))^2 (\mathscr{S}(\omega))^2 d\omega\ \right>,
\end{equation}
where $\mathscr{D}(\omega)^2= P(\omega)$, and $\omega'$ is the angular frequency over which $\mathscr{D}(\omega)$ and  $\mathscr{S}(\omega)$ are defined.  Equation (\ref{eqn:atmospower}) covers a significant portion of the optical spectrum, whereas we are interested in the $z$ band, where the atmosphere is more transparent.  We assume that $P(\omega)$ at the $z$-band scales linearly with the \cite{1994SoPh..152..351H} power spectrum and we derive a scalar coefficient based on our atmosphere model discussed previously and information gathered from the CFHT SkyProbe \citep{2009PASP..121..295S}.  Our atmosphere model gives us the average atmospheric transparency for SDSS~$z$ (or any other bandpass) to compare with \cite{1994SoPh..152..351H}.  We use the CFHT SkyProbe data to derive a scale factor between transparency fluctuations of \cite{1994SoPh..152..351H} and that of the $V$-band.  This is consistent with the results of our atmospheric model; i.e., the amplitude of transparency fluctuations scales with average transparency fluctuations, thus enabling us to properly adjust equation (\ref{eqn:atmospower}) to any bandpass within the range of our atmosphere model.  For our observations, the median noise due to transparency fluctuations is $2.8\times10^{-4}$ and is typically smaller than Poisson and scintillation noise.  

\subsection{Detector Noise}\label{sec:detectorbeh}
\begin{figure}
\centering
\includegraphics[width=\textwidth]{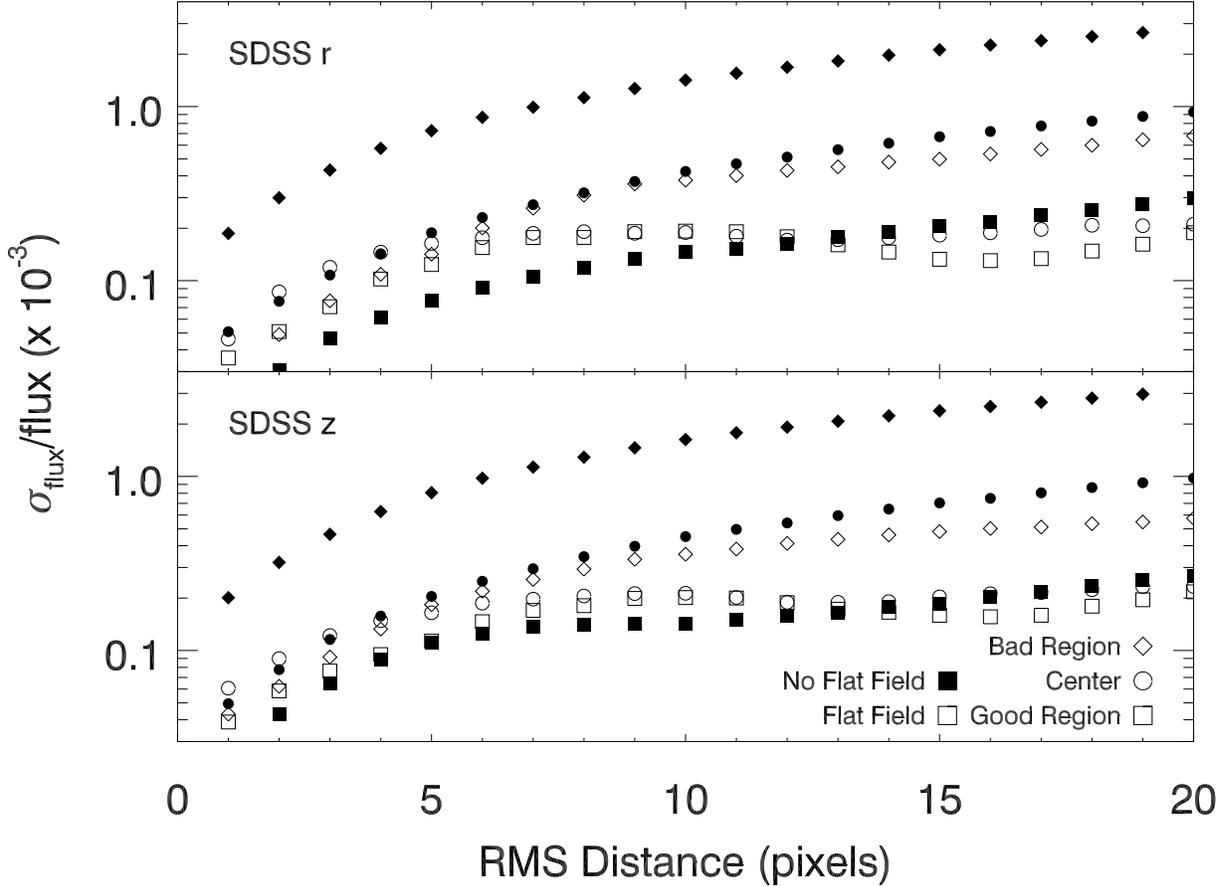}
\caption{Fractional error as a function of rms image motion within a bad region (high pixel response variation), a good region (low pixel response variation), and at the center of the SNIFS science-grade imaging detector.  These curves were generated by taking a defocused image and shifting it around a map of the detector response produced from a series of $\sim$225 flats taken with varying exposure times through the SDSS $r$ (top) and SDSS $z$ (bottom) filters.  Nonfilled points are for a pixel response map with a flat-field correction (median of 10 twilight flats).  We assume no changes within the defocused PSF except position on the detector.  Note that the good/bad regions used are comprised of different sets of pixels for the two different filters used.  
   \label{fig:chipmotion}}
\end{figure}

No CCD has a perfectly uniform response rate, even when corrected with flat-fields.  We calculated the error due to pixel response nonuniformities combined with motion of the defocused image around a given coordinate centroid, ignoring changes within the defocused PSF and chip nonlinearities.  To accomplish this, we took 225 dome flats in 2010 June with exposure times from 2 to 15~s; 150 flats were taken using the SDSS $z$ filter, and the remaining 75 were taken with SDSS $r$.  All flat-fields were obtained within a period of 5~hr.  Count levels varied from $5\times10^{3}$ to $3.8\times10^{4}$~e$^{-}$ (typical levels for the signal in one of our defocused images).   For each filter, we performed a linear least-squares fit of electron counts in each pixel versus median electron counts (for the chip of interest), assuming Poisson and read-noise variance.  The result was a pixel-by-pixel map of the CCD's response for each filter.  The median count was used instead of exposure time to remove effects of lamp variation.  We created a model defocused PSF and scanned it over the $1024^2$ region of interest in our detector response map to find the region(s) of the CCD with the most uniform pixel response.  In general, the CCD's behavior was similar in both filters, although variation of the pixel response for the best (most uniform) regions was slightly better when using the $r$ filter (rms pixel response $\sim 0.7\%$) than it was when using the $z$ filter (rms $\sim 0.8\%$).  For either filter, even small PSF motions ($<$ 3 pixels) on a bad region (very nonuniform pixel response) can cause noticeable ($> 3\times10^{-4}$) variations in received flux.  Large motions ($>$ 15 pixels) around a well-behaved region (highly uniform pixel response) can create similarly large variations in flux.  Because we use a good region on the chip, typical rms motions for our observations ($<$ 5 pixels) contribute minimal noise ($\lesssim1\times10^{-4}$ for $z$ and $\lesssim0.8\times10^{-4}$ for $r$) to the total error budget (Fig.~\ref{fig:chipmotion}).

Pixel-to-pixel variations in the response of the detector can be removed to a certain extent with dome and/or sky flats.  However, using flats introduces additional noise to the data, because quantum efficiency variations among the CCD pixels are wavelength-dependent, the dome lamp (or sky) will not have the same spectrum as the target, and master flats composed of numerous, high-S/N flats still have noticeable Poisson noise.  Twilight flats will better match the spectral energy background distribution of the data frames; however, it is difficult to get a large number of high-S/N twilight flats in the relatively short twilight window.  These and many other errors associated with flat-fielding are discussed more thoroughly in (among others) \citet{1991PASP..103..122N}, \citet{1993spct.conf..304T}, \citet{1995A&AS..113..587M}, and \citet{2001ASPC..238..373M}.  

Noises brought about by flat-fielding may be significant, compared with the $\lesssim 10^{-4}$ sized errors induced from inaccurate detector response (Fig.~\ref{fig:chipmotion}).  For example, consider a master flat composed of 10 twilight flats, each with $\sim 20000$~e$^{-}$ pixel$^{-1}$, and defocused target/comparison stars spread over 1000 pixels.  Taking into account Poisson noise only, the error associated with flat-fielding to the light curve will be $\sim 1\times10^{-4}$, putting it on par with other noise from detector nonuniformity.  Further, when we apply a median of 10 twilight flats to our dense grid of 150 flats (we assume the grid of flats to be a more accurate map of the detector) the predicted errors created by image motion do not improve significantly, as shown in Fig.~\ref{fig:chipmotion}.  Applying a flat-field correction improves precision when using a mediocre region of the chip, and/or when the image is drifting $\gg$10 pixels.  Since we make use of a very flat region of the chip, and image motions are small, flat-field corrections may actually {\it add} noise. 

Detector nonlinearities are small for most science-grade instruments, but they can become important for photometry at submillimagnitude precision.  For a single observation the recorded flux will be approximately
\begin{equation}
S \approx \displaystyle\sum\limits_{i}(x_i-\alpha x_i^{2}),
\end{equation}
where $x_i$ is the normalized incident intensity on any given pixel, and $\alpha$ is the level of nonlinearity in the pertinent range.  If the incident flux is $\bar{x} + r$, where $r$ is normally distributed with standard deviation $\sigma$, then the difference between the signal of any two observations, $S_j$ and $S_k$, will be
\begin{equation}
S_j - S_k  \approx \displaystyle\sum\limits_{i}[2\alpha(r_{ij} - r_{ik})   +  \alpha(r_{ij}^2 - r_{ik}^2)].
\end{equation}
The distribution of $S_j - S_k$ will have an average of zero, but a nonzero standard deviation.  If we ignore terms of higher order than $\sigma^2$, then the latter will be:
\begin{equation}
rms \sim \frac{\sqrt{2}\alpha\sigma^2}{\sqrt{N}},
\end{equation}
where N is the total number of pixels.  A typical aperture size for our observations contains $N \sim$1.5$\times10^4$ pixels.  $\sigma$ can vary significantly between observations, but is no higher than $1.0$ and is often less than 0.1.  In 2005 a series of dome flats were taken to determine the flux range over which the SNIFS E2V CCD is linear, which we use to estimate $\alpha$.  Exposure times varied between 2 and 55~s and were interleaved to reduce any time-varying effects.  All exposures were taken within a few hours.  The region containing the lower 200 pixels of the photometric chip and relevant amplifier was found to be linear to $0.15\% \pm 0.08\%$ between $7\times10^{3}$ and 1$\times10^5$~e$^{-}$.  The rms of these measurements is $0.16\%$, most of which is due to variations in lamp brightness.  Since we require the chip to be linear to better than $0.1\%$ for submillimagnitude photometry, we aim to keep the maximum flux in any given pixel below $5.0\times10^{4}$~e$^{-}$, where the CCD is linear to better than $10^{-3}$.  Thus for flux levels consistent with the majority of our observations, $\alpha \lesssim 0.0015$.  Using $1.0$ as an upper limit on $\sigma$, the noise due to nonlinearities is only $1.7\times10^{-5}$.

\begin{figure}
   \centering
\includegraphics[width=5.7cm]{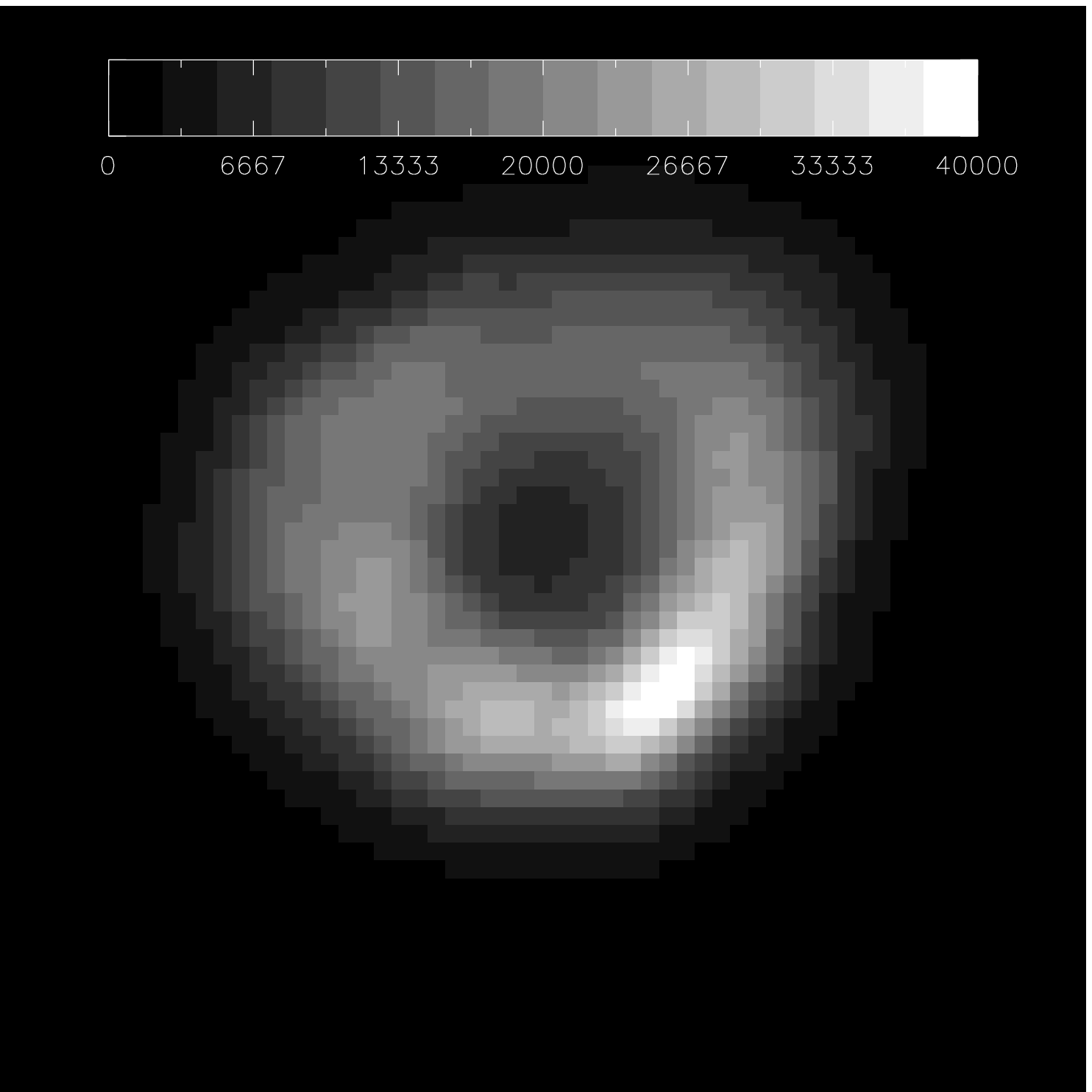}          
\includegraphics[width=7.0cm]{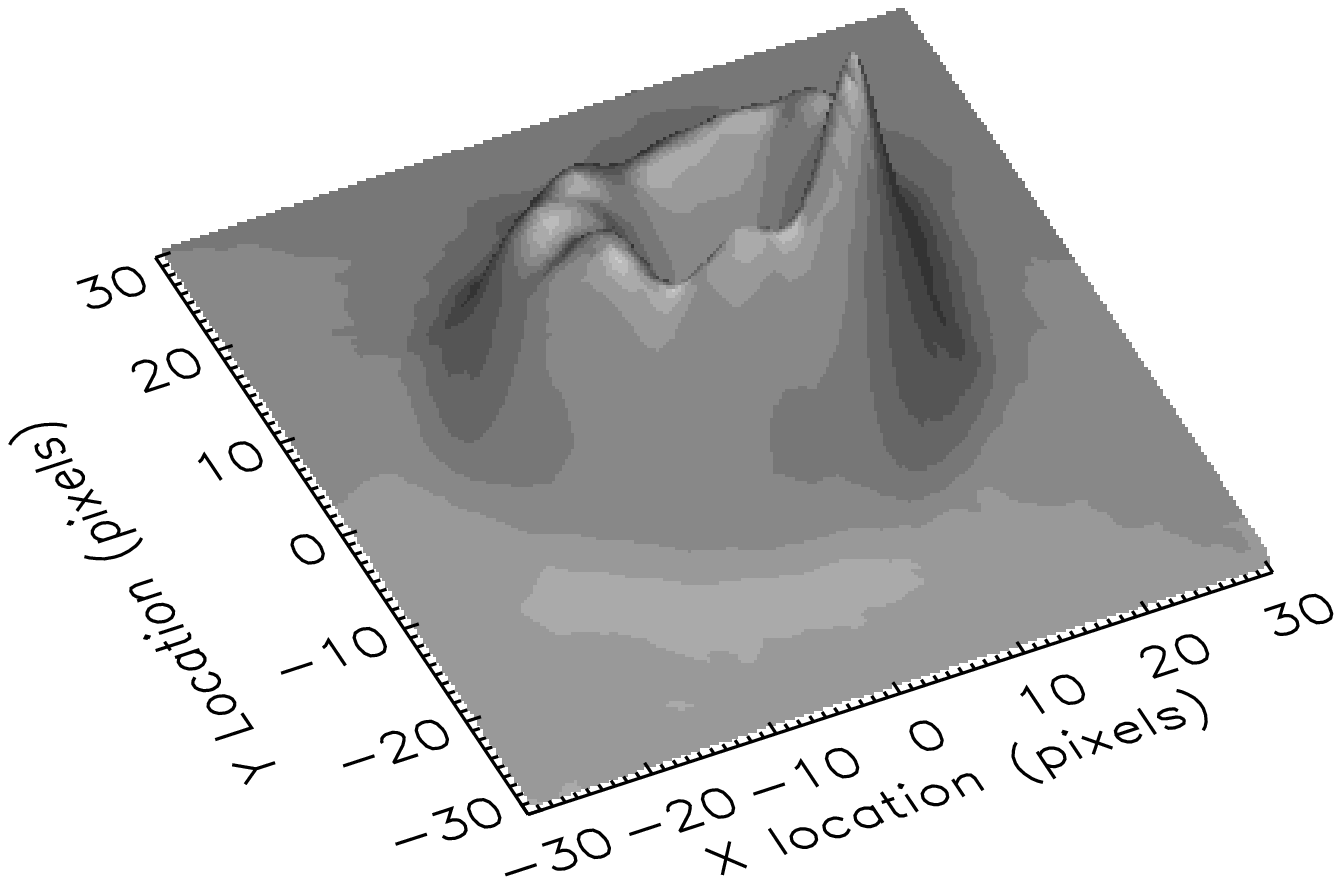}
\includegraphics[width=5.7cm]{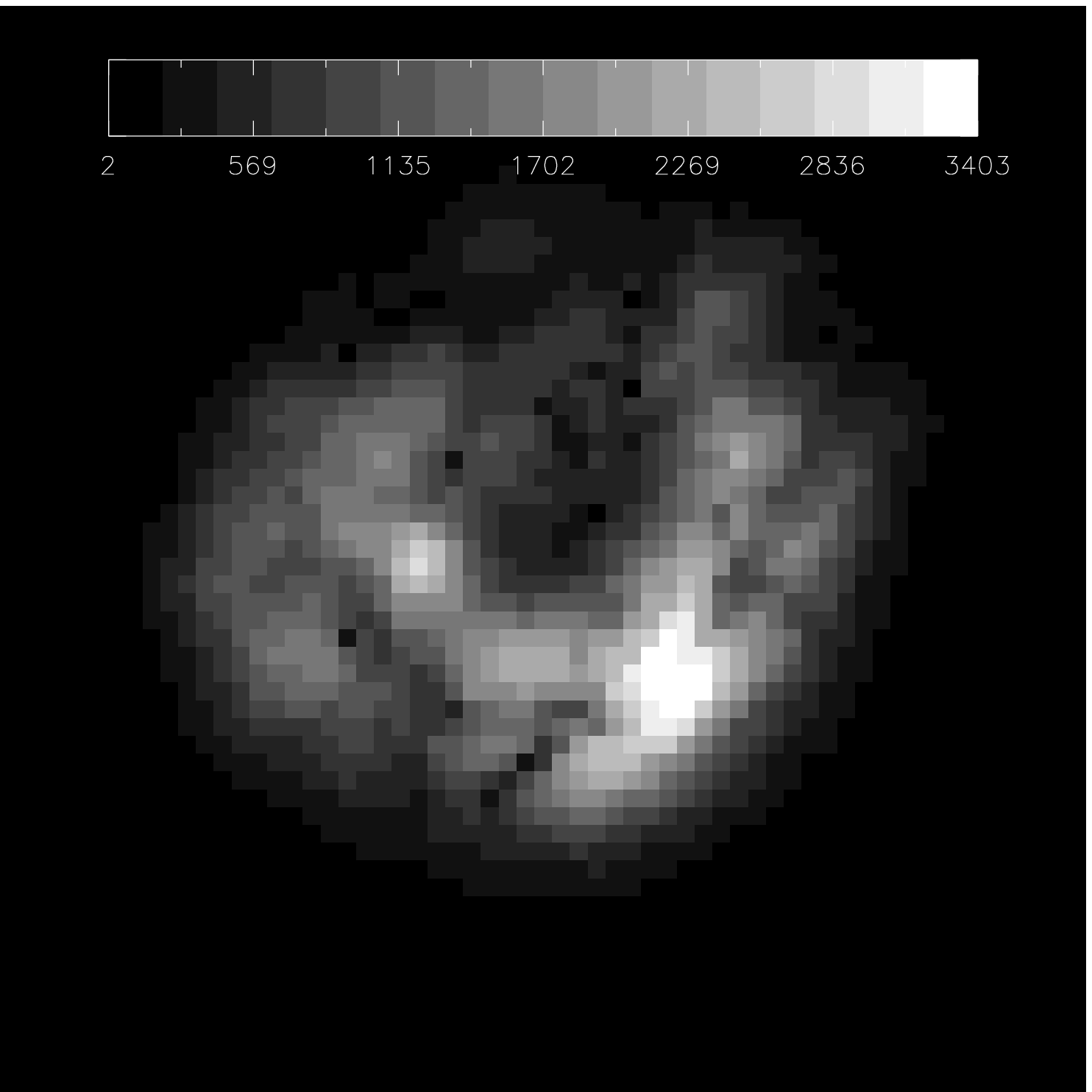}
\includegraphics[width=7.0cm]{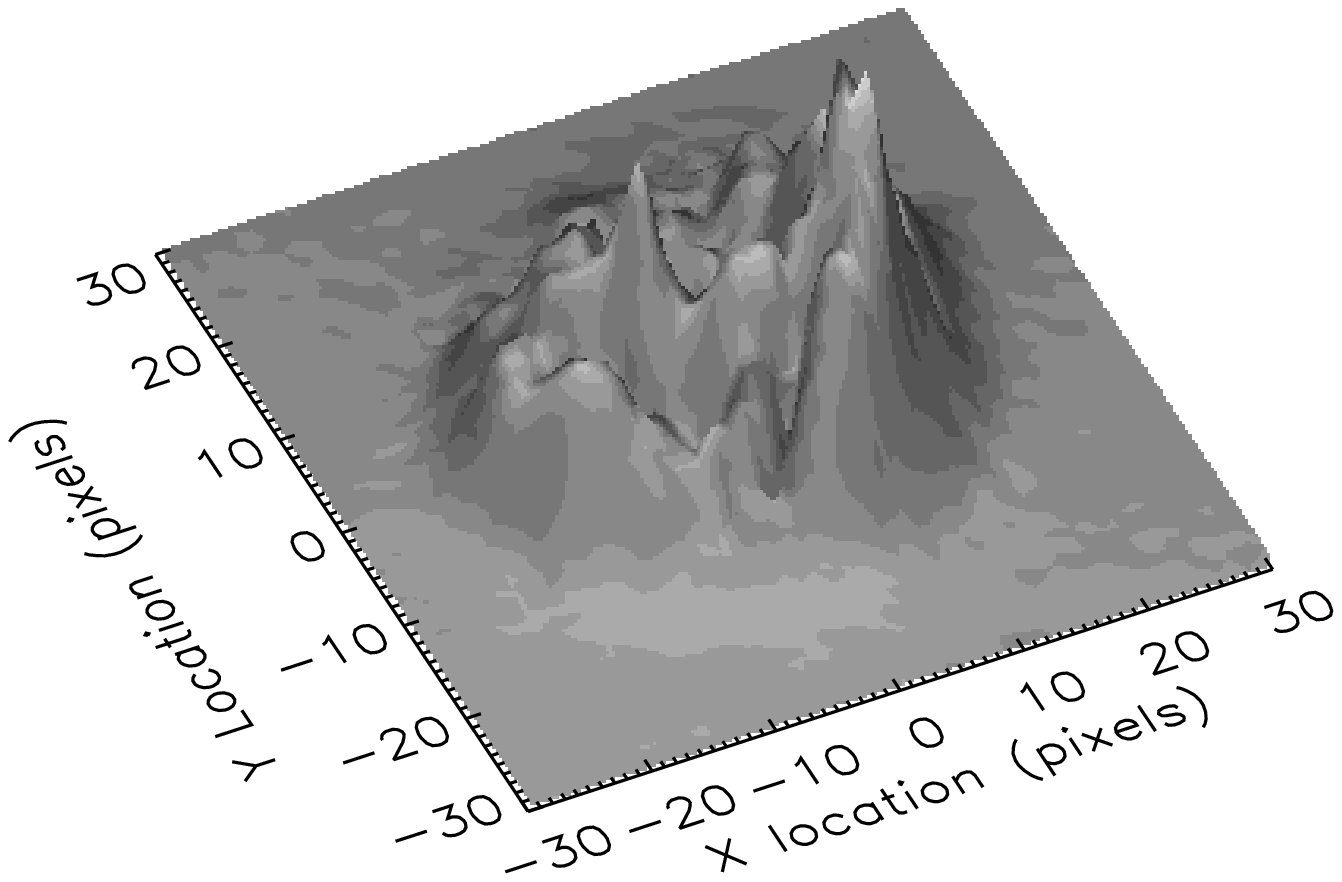}
\caption{{\it Top:} Mean of eight defocused images of a $V=13$ star, each with an integration time of 30~s. {\it Bottom:} Standard deviation of pixel values from the mean.  All images have the same spatial scale ($\sim8\arcsec$ on a side), but not the same intensity scale.  The highest peak in the PSF is due to coma in the UH~2.2~m optics \citep[C. Aspin, private communication 2010;][]{1973A&A....28..355B}. \label{fig:PSF}}
\end{figure}

Another potential source of error is variation in the {\it shape} of the PSF, combined with fixed calibration errors in the pixel response.  Even with sufficient defocusing, phase errors from seeing will induce fluctuations across the face of the image.  The result is that, although total flux may be conserved, flux will be redistributed between pixels, which will produce uncertainties to the extent that pixel responses are not perfectly calibrated.  If the PSF did not vary, such calibration uncertainties would not produce time variations.  Fig.~\ref{fig:PSF} shows the rms difference between eight individual images and the mean.  The stellar wings are mostly constant, but there is significant variability in the shape of the core annulus, particularly at its inner and outer edges.  We estimate the magnitude of such an error by assuming that the calibration suffers from errors having a fixed Gaussian distribution with $\sigma$ = 1\%.  The error calculated from the eight images is only $4\times10^{-5}$, which is a consequence of summing uncorrelated errors over the more than 10$^4$ pixels with which the signal is acquired.

\section{Strategies to Minimize Noise}\label{sec:minimization}
\subsection{CCD Regions and Signal Levels}\label{sec:ccdmitigation}
Large pixel nonlinearities, like those seen in some near-infrared detectors, can be corrected by applying fits to an intensity series for each pixel and then implementing the corrections to the detected flux \citep{2004PASP..116..352V}.  Pixel nonlinearities can also be mitigated in differential photometry by using the same pixels and similar flux levels and exposure times.  More subtle nonlinearities and time-variable pixel responses are difficult to independently measure and remove, but cannot be ignored when submillimagnitude precision is required.   One solution is to identify those pixels that are poorly behaved (highly nonlinear and/or time-variable pixels) and avoid them.  It is {\it not} possible to simply mask these pixels during image processing, because that sensitizes the total signal to small changes in the position or shape of the PSF.  Our strategy for dealing with noise related to the detector is threefold, we locate the range of signal over which the detector is sufficiently linear (discussed in \S~\ref{sec:detectorbeh}), identify regions of the detector containing pixels with uniform response rates, and minimize the motion of the defocused image between integrations.  

\begin{figure}
   \centering
\includegraphics[width=\textwidth]{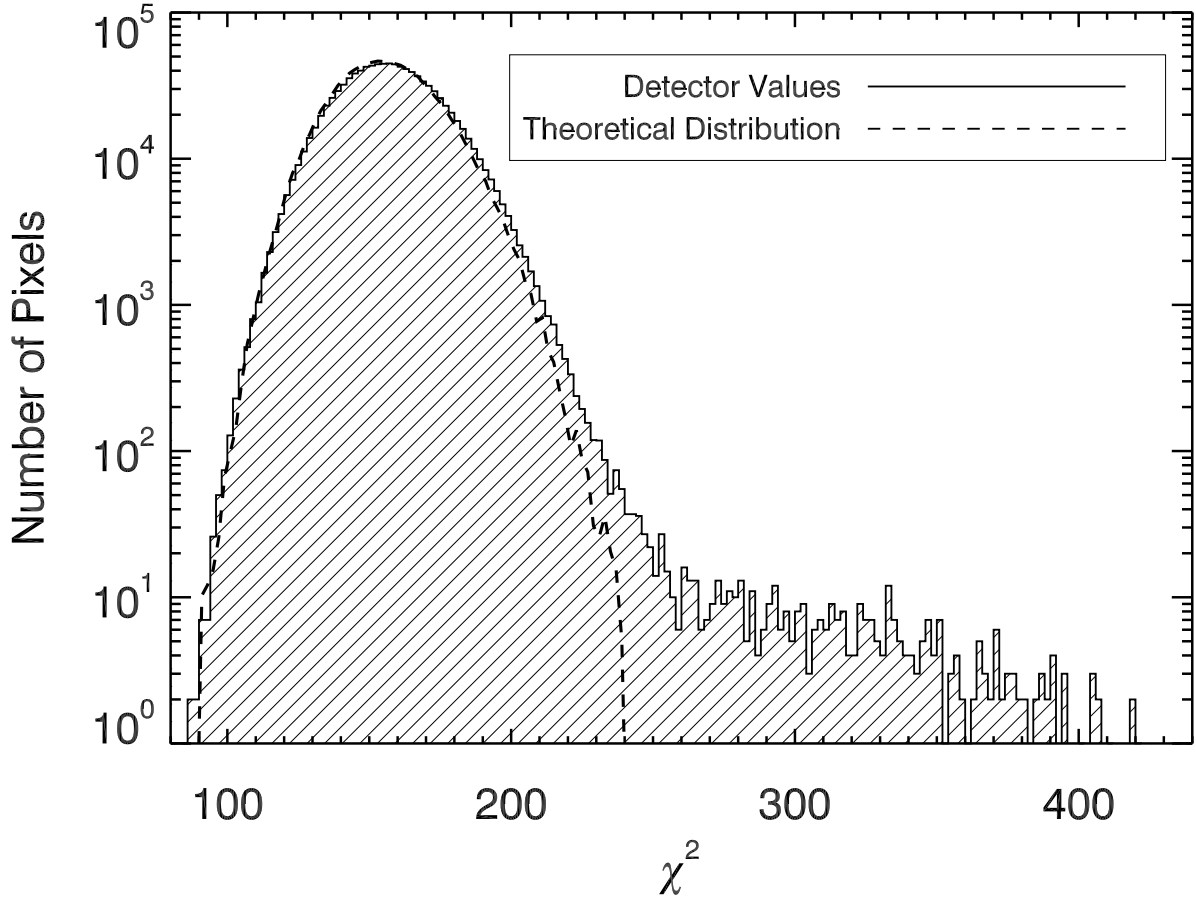}
\caption{Histogram of the $\chi^2$ values of a fit to pairs of individual pixel
  values vs. median detector values in a series of 150 dome flats.  The theoretical distribution derived from a Monte Carlo analysis (based on Poisson statistics) is shown as a dashed line.  We only include data from the bottom half of the SNIFS imaging science-grade detector (the region we use for transit imaging).  We consider pixels with $\chi^{2} > 240$ as poorly behaved and worth avoiding.  These represent $\lesssim 0.1\%$ of the total number of pixels. \label{fig:singlevar}}
\end{figure}

We used the series of flats taken in 2010 (discussed in \S~\ref{sec:detectorbeh}) to locate time-variable or highly nonlinear pixels on the SNIFS E2V detector.  We identified pixels that have the highest $\chi^2$ values for a linear fit.  Fig.~\ref{fig:singlevar} shows a histogram of the $\chi^2$ values for pixel across the $1024^2$ region of the detector used for photometry.  Because our flat experiment is designed to locate the best and worst regions on the detector the absolute variation from linearity is not as important as the identification of pixels that show higher levels of nonlinearity (higher $\chi^2$) than others.  $\lesssim$0.1\% of pixels have $\chi^2$ values above the expected distribution ($\chi^2 > 240$) and are predominantely concentrated in a few areas (e.g., near the edges of the detector), making poorly behaved pixels easier to avoid.  

\begin{figure}
\centering
\includegraphics[width=\textwidth]{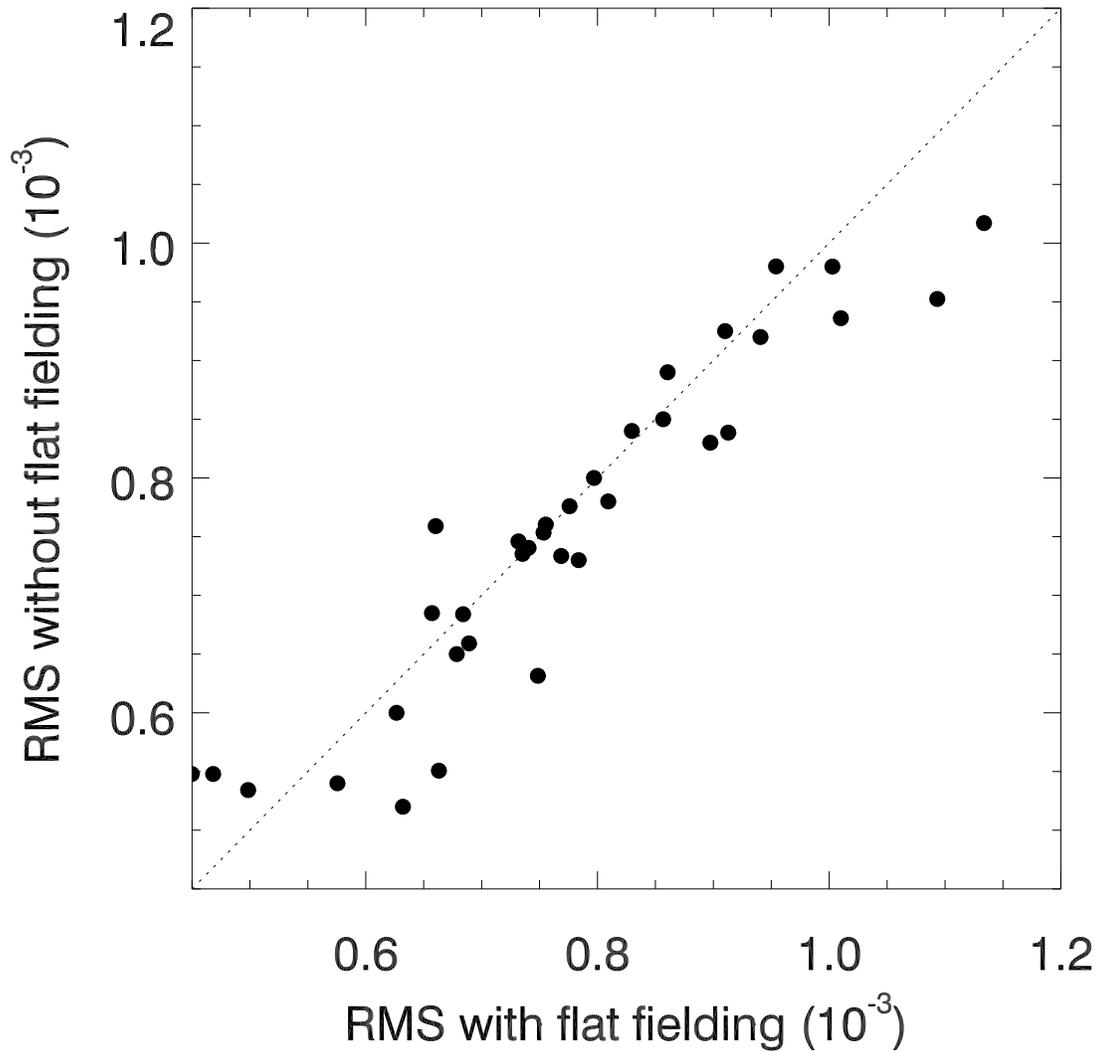}
\caption{Comparison of photometric precision with and without flat-fielding.  Dome flats are used when there is an insufficient quantity of high-S/N twilight flats from a given night.  Although in some cases the precision is better with flats, on average, the precision is better without flat-fielding. \label{fig:flatvsnoflat}}
\end{figure}

Our data indicate that the best precision is achieved {\it without} flat-fielding, provided that images are placed in good (flat) regions of the chip and that image motion is small (rms $<$ 10 pixels).  Fig.~\ref{fig:flatvsnoflat} shows a comparison of precision with and without flat-fielding.  For light curves built with flat-fielding we use twilight flats where available, and dome flats elsewhere.  Although some light curves benefit from the use of flat-fields, applying flat-field corrections to most of our data worsened the precision.  Instead, noise from errors in pixel response are minimized by using the same (good) region of the chip for all observations with a given filter (see Fig.~\ref{fig:chipmotion} for a comparison of noise from good and bad regions of the CCD) and by keeping image motion small from integration to integration.  SNIFS' automated acquisition and position of the target/comparison stars is more than sufficient to keep image motion below 5~pixels.  Flat-fielding will have more beneficial results for detectors with higher interpixel QE variations or higher image drift between integrations.

\subsection{Choice of Filter}
\begin{figure}
   \centering
\includegraphics[width=\textwidth]{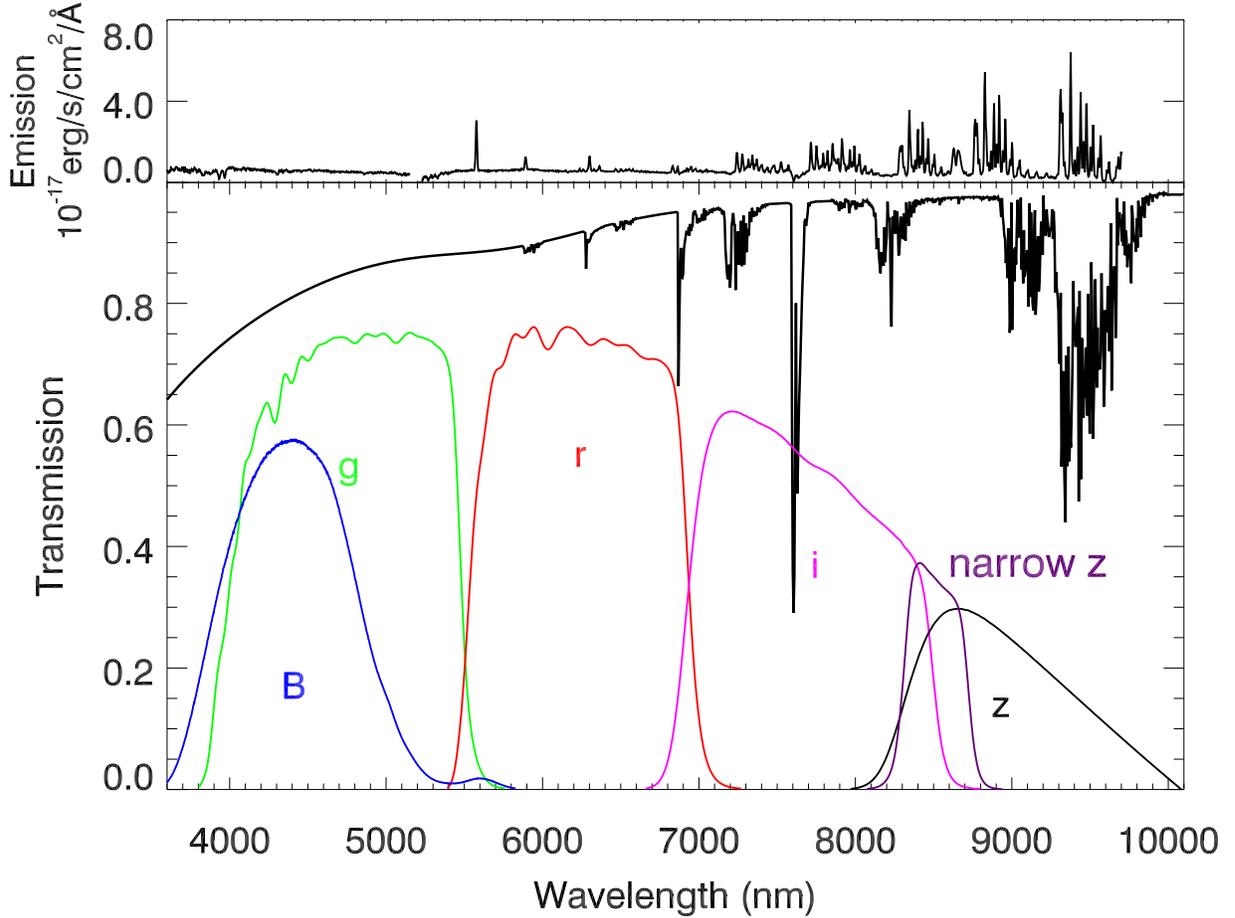}
\caption{Approximate transmission (bottom) and emission (top) of the atmosphere on a clear night over Mauna Kea with $B$, SDSS~$g$, SDSS~$r$, SDSS~$i$, SDSS~$z$, and narrow $z$ filters multiplied by the SNIFS CCD quantum efficiency.  There is a discontinuity at $\sim5200$~\AA~ in the emission caused by low QE near the edge of the blue and red channels of the SNIFS integral field unit.  Our observations were taken predominantly in SDSS~$z$, although some were taken in SDSS~$r$.  Scattering affects the filters blueward of 5500~\AA~whereas molecular (chiefly \water~and O$_2$) absorption lines contaminate the $r$ and $z$ passbands.  A narrow $z$ filter (J. Johnson, private communication) can mitigate errors from both molecular absorption and scattering.  \label{fig:atmosfilters}}
\end{figure}

Choice of bandpass filter is a tradeoff between the desire to maximize signal from the star and to minimize the contribution from wavelengths that are affected by scintillation, scattering (by aerosols and molecules), and absorption by \oxygen~and \water~\citep{2007PASP..119.1163S}.  Observations through redder filters will have lower first- {\it and} second-order extinction (see \S \ref{sec:extinction}).  Stellar variability is a significant source of noise in the $V$ band for as much as $20\%$ of M dwarfs on 12~hr timescales \citep{2011AJ....141..166H, 2011AJ....141..108C}.  However, observing at longer wavelengths reduces the spot-photosphere contrast, reducing noise from stellar rotation and spots.  Among the redder SDSS filters, $i$ is seriously compromised by molecular bands (Fig.~\ref{fig:atmosfilters}), $z$ is bracketed by two \water~bands, although the low quantum efficiency of most CCDs redward of 9000 \AA~mitigates \water~contamination, while $r$ is more affected by aerosol fluctuations.  Use of a narrow $z$ filter (if one is available) would almost completely eliminate second-order extinction error (Fig.~\ref{fig:specerror}), due to its narrow bandwidth and red central wavelength (Fig.~\ref{fig:atmosfilters}).  A narrow $z$ filter would also reduce noise from scintillation and scattering from aerosols while avoiding a major \water~line.

Observations in near-infrared filters are more affected by atmospheric emission than bluer bands, however, our data suggest that the overall noise from atmospheric emission and/or fringing is small.  Fringing is present in SDSS~$i$ and $z$ for SNIFS, but noise from fringing is minor compared with the rest of the sky background.  Since we predominantly observe within 7 days of a full Moon, the dominant component of the sky background is scattered light from the Moon.  Fig.~\ref{fig:atmosfilters} includes an estimate of the atmospheric emission over Mauna Kea based on spectra taken from SNIFS.  In the $z$ band, integrated sky emission in a $\sim 7~\arcsec$ radius, a typical defocus size, will be at least a factor of 100 smaller than the flux from a typical star ($m_z$ = 9) in our program.  In the narrow $z$ band, this grows to 180, since the narrow $z$-filter transmission cuts off before the largest OH emission lines that contaminate the $z$ band.  

To test the performance of SDSS~$z$ and $r$ filters we performed regular observations while interleaving exposures of $r$ and $z$ filters.  We kept other controllable variables (exposure time, defocus, etc.) fixed between filter changes.   The $z$ band slightly outperformed the $r$ filter overall (see Table \ref{tab:table2}), but not by as much as our total noise estimate predicts (see \S~\ref{sec:noisemodel} for more on the noise model).  The discrepancy could be due to underestimating the effect of the \water~line on the $z$ filter, which is dependent on the PWV of that particular night and induces noise that cannot be easily modeled.

\subsection{Degree of Defocusing}\label{sec:defocuPSF}
A minimum degree of defocusing is required for a fixed total signal from the star and linear range of the detector.  For example, if the linear range of a CCD is $5\times10^4$~e$^{-}$, then the acquisition of $2\times10^7$~e$^{-}$ must take place over at least $400$ pixels (for SNIFS, this is a circle 5$\arcsec$ across).  Because the PSF is nonuniform, more pixels must actually be used to collect the signal and avoid saturation.  Greater defocusing may be desirable for very bright stars to avoid integration times much shorter than the CCD read time and to minimize scintillation noise.
  
\begin{figure}
\centering
\includegraphics[width=8.0cm]{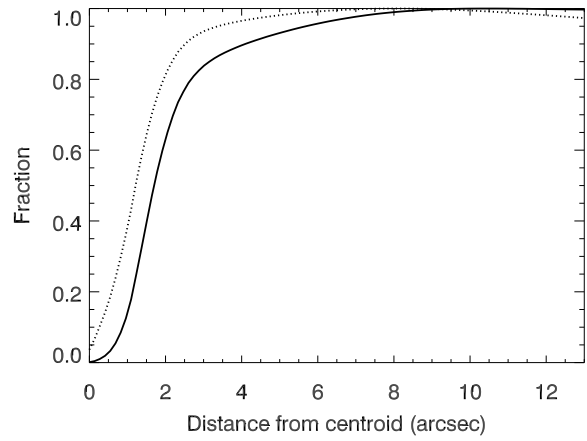}
\caption{Curve of growth (vs. radius in arcseconds) of the mean of eight defocused images of a star obtained with the SNIFS/UH~2.2~m (solid line), and the S/N in an aperture of a given radius relative to the maximum value (dotted line). \label{fig:cog}}
\end{figure}

Fig.~\ref{fig:PSF} is the mean of eight 30~s integrations of a defocused $V\approx13$ M star obtained through the SDSS~$z$ filter.  Most of the signal is confined to an annulus with a $\sim$4$\arcsec$ radius that represents an out-of-focus image of the telescope's primary mirror. The defocused PSF core is surrounded by a halo pattern, produced by the defocused convolution of the telescope pupil with the seeing, which extends out $\sim$10\arcsec.  The image is not axisymmetric because of coma in the telescope optics \citep[C. Aspin, private communication 2010;][]{1973A&A....28..355B}.  The area within the annulus is $\sim$10$^3$ pixels, and the average signal per pixel is $1.7\times10^4$~e$^{-}$, with a total signal of 1.6$\times 10^7$~e$^{-}$.  The total read and electronic pickup noise (200~e$^{-}$) is much less than the photon noise ($4\times10^{3} $e$^{-}$).  The maximum S/N is achieved by summing the signal within an aperture of 8\arcsec radius, if only photon noise, sky noise, read noise, and electronic pickup noise are considered (Fig.~\ref{fig:cog}).

More defocused images are less sensitive to motion of the stellar image on the CCD; however, the total read noise and probability of including ill-behaved pixels are greater.  We consider the idealized case where the PSF is a circular aperture comprising $N$ pixels, each of which gets $S\times N^{-1}$ signal (e$^{-}$) and has read and electronic pickup noise $\sigma_{1}$ (e$^{-}$ pixel$^{-1}$) and uncorrected response variation $\sigma_2$, which we presume adds in quadrature.  The variance from a single nonoverlapping pixel due to the last source of error is $S^2N^{-2}\sigma_2^2$.  If the centroid rms motion in pixels is $\delta$ then the number of pixels that are not common to a pair of stellar images is $\sim$2$\pi \delta \sqrt{N/\pi}$.  The total variance due to read noise and image jitter is
\begin{equation}
\sigma^2 \approx N\sigma_1^2 + 2N^{-3/2}\sqrt{\pi}\delta S^2\sigma_2^2.
\end{equation}
Total read noise increases with $N$, but noise due to image jitter decreases.  The number of pixels $N_*$ that minimizes the total error is
\begin{equation}
N_* \approx \left(2 \delta \sigma_2^2 S^2\sigma_1^{-2}\right)^{2/5}.  \label{eqn:nstar}
\end{equation}
For $\delta = 3$ (the median centroid motion in pixels for our observations), $\sigma_2 = 0.01$, $\sigma_1 = 6 $e$^{-}$, and $S = 10^7$e$^{-}$, and $N_* \approx 7.2\times10^3$: i.e., a circular region with diameter of $\sim$96 pixels ($13~\arcsec$) with total noise $\sigma = 6.6\times10^{-5}$~e$^{-}$.  $N_*$ is sensitive to well depth, read noise, and the uncorrected pixel response noise, but can be estimated by experiments with defocused images.  We are often forced to use higher defocus values than equation (\ref{eqn:nstar}) suggests to keep the counts in the coma-induced peak in the linear range of the detector.  

\subsection{Integration time}\label{sec:inttime}
\begin{figure}
\centering
\includegraphics[width=\textwidth]{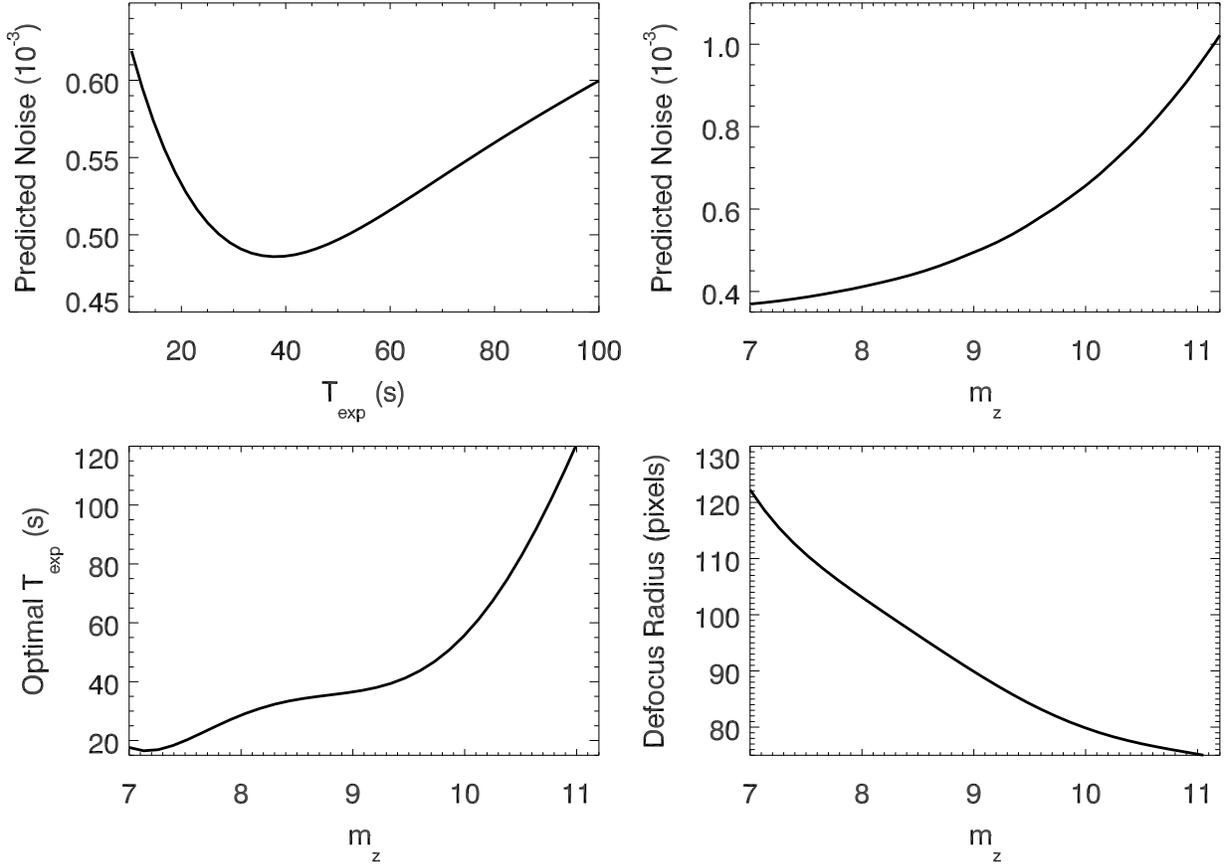}
\caption{Total predicted noise as a function of exposure time ({\it top left}) for a $m_z$ = 9 star.  Also shown is the total noise ({\it top right}), optimal defocus radius ({\it bottom right}), and exposure time ({\it bottom left}), corresponding to the setup with the lowest theoretical noise as a function of $z$ magnitude.  We perform these calculations based on noise from Poisson statistics, read and electronic pickup noise, scintillation, sky background, and atmospheric transparency fluctuations.  We have assumed 1.5 minutes for readout and slew time between stars and that the comparison and target star are of identical flux and spectral type.  Note that for our stars (mostly late K and early M dwarfs) $V-z \sim$1.8 \citep{2007AJ....134.2398C}.}\label{fig:exptimetests}
\end{figure}

Poisson and scintillation noise decrease with integration time whereas noise from ATF will increase.  Additionally, higher exposure times for a given star will require more defocusing, increasing contributions from read and sky noise.  There exists an optimal exposure time that minimizes the total error from these sources.  We estimate the optimal exposure time as a function of the magnitude of the star using our calculations for all of these sources of noise: ATF, scintillation, Poisson, read and electronic pickup, and sky.  We use a model of our defocused images (including the coma-induced peak) to calculate the required defocus in order to satisfy equation (\ref{eqn:nstar}) and to keep the flux below $5\times10^4$~e$^{-}/$pixel (i.e., where the detector is linear to better than 0.001).  We then calculate the total expected noise as a function of exposure time, assuming target and reference stars of identical flux and spectral type, yielding an optimal exposure time for a given stellar magnitude.  We repeat this calculation for a range of magnitudes encompassing those in our survey (see \S~\ref{sec:obs}).  Fig.~\ref{fig:exptimetests} shows the optimal exposure times, defocused image radius, and calculated noise for the exposure time with the best expected precision.  Our calculation overestimates exposure times for the brightest stars ($m_z \lesssim  8$), where the suggested level of defocusing becomes impractical.

In our exposure time tests, we interleave observations of a pair of stars ($m_z = 10.8$ and $11.4$) using three sets of exposure times each (25s, 50s, 100s and 40s, 80s, 160~s, respectively).  According to our applied noise model, the best precision should be obtained with the 50s and 80s integration times (Fig.~\ref{fig:exptimetests}).  However, the lowest total noise came from the highest exposure time (100s and 160s).  Additionally, the 50/80s and the 100/160s exposure time sets beat the expected precision (see Table \ref{tab:table2}), suggesting that ATF may have been lower on that particular night, or that we might be averaging over stellar P-mode oscillations with longer exposure times \citep{1991PASP..103..221Y, 1994ARA&A..32...37B, 1999PASP..111..845H,2011AJ....141..108C}.  

\subsection{Choice of Comparison Star}
There are noticeable gains from proper selection of a comparison star.  Our collection of low-resolution spectra enables us to select comparison stars of a similar spectral type to our target, resulting in very small errors from second-order extinction ($4.5\times10^{-5}$).  If we replace all comparison stars with G dwarfs (6500~K), the median value becomes $3.0\times10^{-4}$, which would make it a more significant part of the error budget (Table \ref{tab:noisetable}).  This is especially important with M-type target stars, which are intrinsically faint and are therefore less likely to be selected as comparison stars based on flux alone.  

\begin{figure}
\centering
  \includegraphics[width=\textwidth]{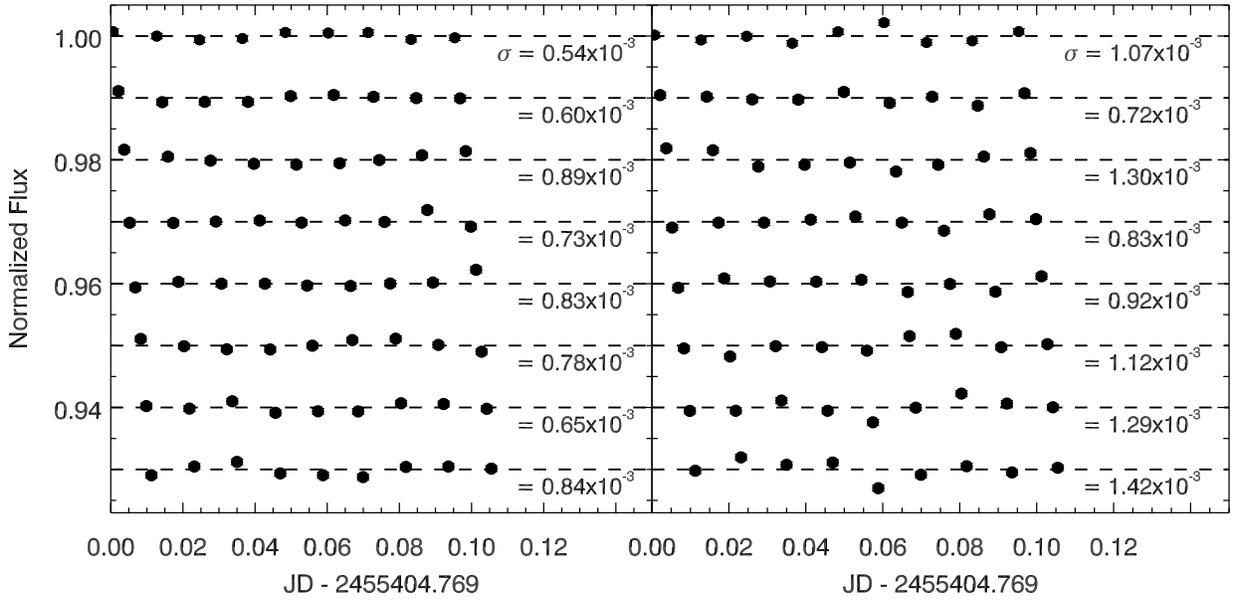}
  \caption{Light curves of eight stars using the constellation method along with rms using a reference signal constructed from stars observed immediately before and after the target (left), and from all seven other target stars (right).  Cadence for these observations is $\sim$ 1 observation every 17~minutes, which is sufficient to sample a transit (transits typically last $\gtrsim$ 1~hr).  All light curves were normalized to 1, and then each was offset in increments of 0.01 for plotting.  Although using more comparison stars improves the S/N, we achieve the best overall photometric precision for each target star using a reference signal built from the stars observed immediately after and before it. \label{fig:constellation}}
\end{figure}

In principle, more comparison star observations will improve the S/N, but these additional observations will occur at increasingly earlier or later times and can engender greater systematic error due to changes in the atmosphere.  We examine the constellation approach by linear construction of a reference signal $\bar{s}$ for the $j$th observation of the $i$th star:
\begin{equation}
\bar{s}_{ij} = \displaystyle\sum\limits_{kl} a_{ijkl}s_{kl},\label{eqn:refsignal}
\end{equation}
where $s_{kl}$ is the flux, and the weighting function $a_{ijkl} = 0$ for $i=k$.  In general, for nonzero values, $a_{ijkl}$ will depend on the characteristics of the atmosphere, the cadence of the observations, and the stability of the data from the stars ($i \ne k$) in the constellation.   To keep the signal normalized, for all $i$, $j$ we force:
\begin{equation}
\displaystyle\sum\limits_{kl} a_{ijkl} = 1.
\end{equation}
We use the constellation technique as an experiment to determine the $a_{ijkl}$ that produces the best photometric precision. We compute a grid of reference signals for each star following equation (\ref{eqn:refsignal}) and apply a range of possible values for  $a_{ijkl}$.  For our observations, we find that the best precision is achieved most consistently when using a reference signal formed from the stars immediately before and after the target observation, i.e., $a_{ijkl} \sim 0$ for all $|i-k| \ne 1$ or $j \ne l$.  This conclusion is consistent with our estimates of the noise from ATF (\S~\ref{sec:transvar}), which begin to overwhelm the noise budget when more than $\sim 7$ minutes passed between target and comparison star observation.  However, the additional stars in the constellation can be used to identify and mitigate noise from aberrant data points (e.g., data contaminated by cosmic rays), or variable stars in the constellation.  Fig.~\ref{fig:constellation} shows the precision for light curves built using the stars before and after the target observation alongside light curves (of the same stars) constructed using an equal weight of all available comparison stars.

\section{Comparison to Observations}\label{sec:noisemodel}
\subsection{Total Errors}
We calculate the total predicted error of each observation by adding the uncorrelated noise from both the target and comparison star(s) due to: scintillation, Poisson (of the star), sky (background), motion on the chip, change in shape of the defocused image, overall chip/amplifier nonlinearities, and readout (read plus electronic pickup noise), along with first- and second-order extinction and ATF.  Noise from motion on the chip is estimated using our model of detector behavior (\S~\ref{sec:detectorbeh}).  The noise from the change in shape (changes in distribution of flux over the defocused image) is calculated assuming that the detector interpixel response suffers from $1\%$ Gaussian variation.  We correct for noise from first-order extinction when this noise source is greater than $1\times10^{-4}$ and we know the extinction coefficient to better than 0.01~mag air mass$^{-1}$.  Although specific levels for each term vary significantly between observations, the median values of Poisson noise are the largest, followed by noise from ATF, scintillation, first-order extinction, sky, read and electronic pickup, motion on the chip, second-order extinction, changes in shape of the PSF, and overall/amplifier nonlinearities (typical noise levels are listed in Table \ref{tab:noisetable}).  The small relative size of many of these errors (second-order extinction, for example) is mostly due to judicious choices in each of our observations: e.g., wise choice of comparison stars, exposure times, and region of the detector used.

\begin{figure}
\centering
  \includegraphics[width=\textwidth]{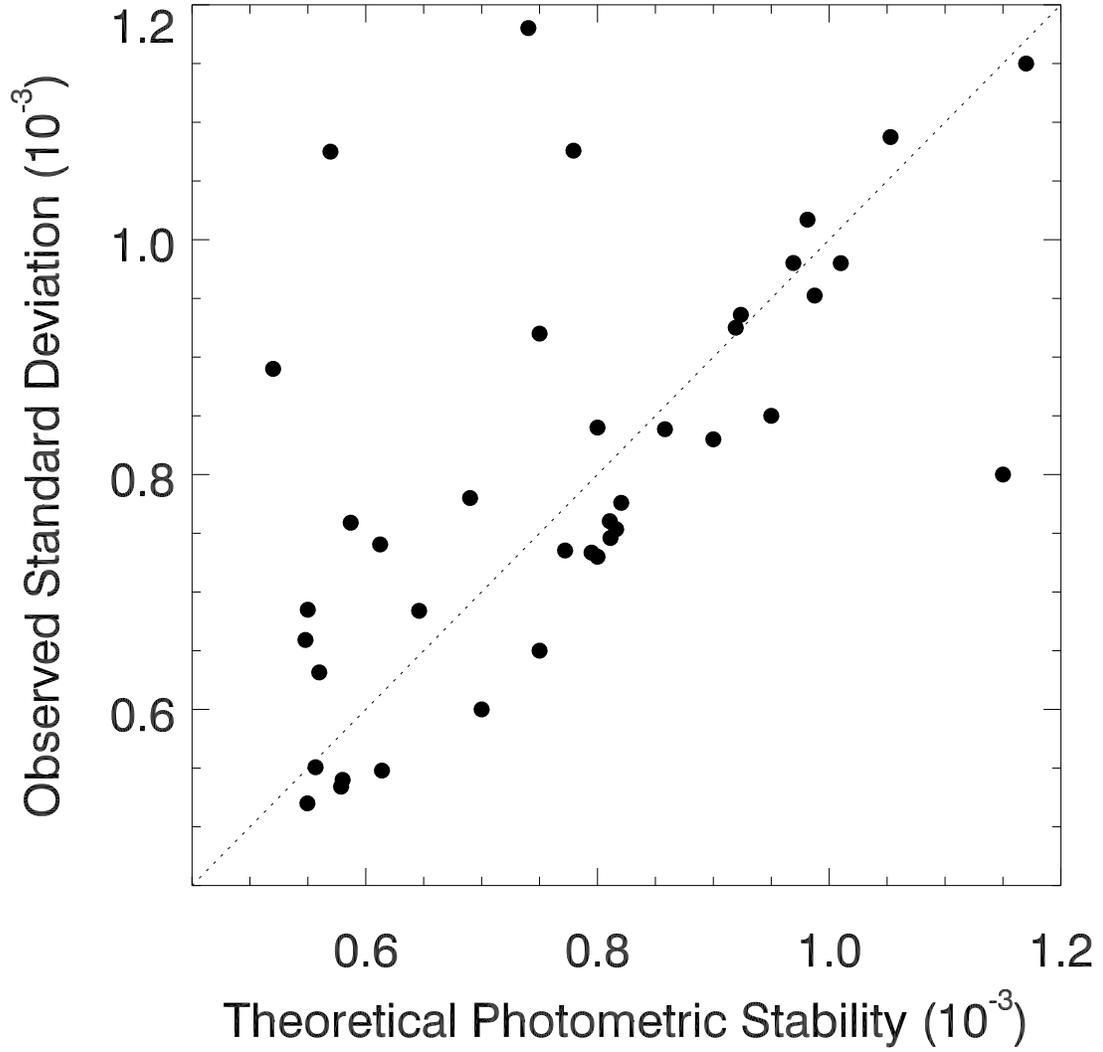}
  \caption{Precision of each of our observations vs. our estimated theoretical precision calculated for each observation.  The dashed line marks where observed precision matches theoretical precision.} \label{fig:theoretical}
\end{figure}

Fig.~\ref{fig:theoretical} shows the precision of each observation versus the total theoretical noise for each observation.  Although the mean of the predicted error is within 3\% of the theoretical precision, there is significant scatter (rms = 12\%).  We expect the major discrepancy to be in our estimate of error from ATF, which is difficult to calculate for any specific observation.  Other noise terms are approximate, such as scintillation, which is only good to $\sim$20\% (even with modification for wind direction).  We have also assumed no correlation between terms, which might be causing us to systematically underestimate noise.  Additionally, stellar variability may contribute to the noise budget, e.g., P-mode oscillations, which can be noticeable for our targets (mostly late K and early M dwarfs) at our level of precision \citep{1994ARA&A..32...37B, 1999PASP..111..845H,2011AJ....141..108C}.  Observations with short exposure times are especially susceptible to P-mode oscillations, as P-mode oscillations have timescales on the order of minutes \citep{1991PASP..103..221Y}.  Thus, our model correctly accounts for the overall magnitude of errors but cannot reliably predict the error of a specific observation.

\subsection{Correlated (Red) Noise}
The performance of most transit surveys is limited largely by time-correlated (red) noise, which limits gains in precision from binning/phasing data \citep{2006MNRAS.373..231P, 2009ApJ...704...51C}.  Red noise is also especially deleterious because it can  mimic a signal of interest, leading to precise but inaccurate results.  We expect to have some degree of time correlation in noise from extinction, ATF, and scintillation, which will manifest as a combination of white and red noise.  Due to all these considerations, we analyze the level of red noise in our data.  We estimate red noise from our comparison-star-corrected light curves using a wavelet-based method.  The method, described in \citet{2009ApJ...704...51C}, determines parameters of noise formed as a combination of Gaussian (white) noise and noise with a power spectral density varying as $1/f^{\gamma}$.  We test the simple case of $\gamma = 1$, as well as the case of $\gamma = 1.5$, implied by equation (\ref{eqn:atmospower}).  No light curve shows red noise above $6\times10^{-4}$ for either $\gamma = 1$ or $\gamma = 1.5$, and median red noise is only $2.8\times10^{-4}$, at least some of which is due to stellar variability.  The low levels of red noise compared with total noise suggest that we are not being overly hindered by time-correlated noise sources.  Thus data acquired using the snapshot technique can be binned to produce improved precision; after binning each observation to a cadence of less than 20~minutes, the median rms from our observations is $3.2\times10^{-4}$ (all over intervals of at least 2~hr to ensure sufficient data points) and all binned light curves had rms of less than 10$^{-3}$.  Further precision improvements can be made with additional binning; however, a cadence of greater than 20~minutes is impractical for detecting the presence of a transit.

\section{Discussion} \label{sec:discussion}
Using defocused snapshot photometry, we have consistently achieved submillimagnitude photometry from the summit of Mauna Kea under clear conditions.  Our lowest photometric rms was $5.2\times10^{-4}$ (with a 5 minute cadence) over a $\sim$2~hr interval.  Of our 38 experiments using the snapshot method, 32 of them had rms precision of less than 1$\times10^{-3}$, and our median precision was $7.8\times10^{-4}$.  We include our constellatio- mode observations, where we observe eight stars in sequence (Fig.~\ref{fig:constellation}).  Although our precision is not as good as the best recorded from the ground \citep[e.g.,][]{1993AJ....106.2441G, 2005AJ....130.2241H}, it is sufficient to detect the transit of a Neptune size planet around a dwarf M0 star or a 10$M_{\oplus}$ super-Earth around an M4 star at 5$\sigma$ significance.  Further, our technique does not rely on a large number of comparison stars, yields submillimagnitude photometric precision consistently, and can be readily automated.  

\subsection{Methods of Submillimagnitude Photometry}
Although our methods were optimized for the UH~2.2m telescope, SNIFS instrument, and Mauna Kea observing site, we give guidelines for achieving submillimagnitude photometry that can be followed by anyone.  Some of these are recapitulated guidelines offered by \citet{1991PASP..103..221Y} and \citet{1999PASP..111..845H}.

{\it Appropriate selection of comparison star:} Optimal choice of comparison star is a tradeoff between minimizing the separation on the sky and using a star of comparable apparent brightness and spectral
type.  Submillimagnitude photometry becomes significantly more difficult with separations of more than a few minutes of slew time between target and comparison stars.  Using more than two comparison stars provides no significant improvement (and usually yields inferior precision), as changes in the atmosphere grow larger than other sources of noise related to the comparison star(s) after just 7 minutes.  Using two comparison stars is preferred over a single star in order to identify any signal that is the result of stellar variability instead of a transit.  Large separations increase air mass differences, and give the atmosphere time (and distance) over which to change.  Although stars of similar apparent brightness are preferred, this technique allows the observer to use different exposure times and focus settings for target and comparison stars.  Thus, significantly dimmer or brighter comparison stars can be used with only minimal loss in precision.  The spectral type of a comparison star should be similar to that of the target, especially if the stars are observed over a wide range of air masses.  It is not usually worthwhile to use comparison stars that require more than $\sim 6$ minutes of time between target and comparison star observation. Past this, the precision loss from atmospheric fluctuations overwhelms noise from a poor choice of comparison star.

{\it Exposure times and defocusing:}  Integration times and the amount of defocus must be chosen so that the signal per pixel is within the linear range ($\ll 1\%$ nonlinear) of the detector.  \citet{1991PASP..103..221Y} suggest using shorter integrations to move rapidly between target and comparison stars.  Our calculations and data indicate that the optimal exposure time depends on the tradeoffs between ATF and shot/scintillation noise, and it often exceeds 1~minute.  Fig.~\ref{fig:exptimetests} shows how to scale exposure times and defocus radius with the brightness of the star, although the values may be significantly different for other telescopes.  Assuming there are no stars that will overlap with the target when defocused, and that the detector has a sufficiently large region of well-behaved pixels, it is best to find the ideal exposure time for each star, and defocus as appropriate to keep flux levels below the nonlinear range of the detector. 

{\it Minimize time between observations:} Making use of a smaller region of the chip will reduce overhead from CCD readout enough that the time between observations is usually limited by telescope/dome slew time.  In the constellation
method, a number of stars are observed in sequence to search for transits (Fig.s \ref{fig:skymap} and \ref{fig:constellation}).  In order to maximize observing efficiency and the number of targets that can be observed with the highest cadence ($>$~1 observation every 30 minutes), a time-efficient trajectory must connect the target stars.  With an efficient path, the constellation technique can be expanded to observe a large number of stars in sequence at the cost of cadence.  If the dome motion is slower than the motion of the telescope, the fastest path minimizes azimuthal moves.  

{\it Observe near meridian crossing:} In addition to being at minimum air mass, objects near meridian crossing experience the smallest changes in air mass relative to nearby stars.  When to observe is especially important when observing over longer periods of time ($>$~1~hr).  For observations on the same set of stars lasting significantly longer than a few hours (producing large changes in the air mass difference), it will be necessary to take enough measurements of the extinction coefficient to constrain it to better than 0.01~mag air mass$^{-1}$ during the observation.

{\it Selection of passband filter:} We achieved similar photometric precision in both SDSS~$r$ and $z$ passbands, possibly a result of higher scintillation noise in the former and contamination by the \water~lines in the latter.  A  narrower $z$ filter lacking the \water-affected wings (Fig.~\ref{fig:atmosfilters}) would avoid both of these problems, as well as reduce the size of second-order extinction.

{\it Use of well-behaved regions of the detector:}  Regions of the detector containing nonlinearities, variable behavior, or significant nonuniformity must be avoided and cannot simply be corrected by flat-fielding (Fig.~\ref{fig:singlevar}).  One significant advantage of the snapshot method is that images of the comparison star and target star can be placed in the same position on the detector.  The rms of centroid motions (from integration to integration) should be kept to less than 10 pixels (Fig.~\ref{fig:chipmotion}).  Snapshot photometry only requires a field of view large enough to fit a single defocused star and enough surrounding field to estimate the sky background with an absolute error much less than that of the star.  

\subsection{Performance of a Hypothetical Transit Search}\label{sec:transitsearch}
To close, we examine the performance of the photometric method described in this article using Monte Carlo simulations of a search for transiting planets around M dwarf stars.  We assume a telescope like the UH~2.2m at a photometric site such as Mauna Kea.  The all-sky input catalog consists of 13,570 nearby late K and early M stars selected from photographic plate surveys based on their proper motions ($\mu > 0.15\arcsec$~yr$^{-1}$)  \citep{2005AJ....129.1483L, 2011AJ....142..138L}, and $V$-$J$ colors, where $J$-band magnitudes come from the Two Micron All Sky Survey, a near-infrared all-sky survey \citep{2006AJ....131.1163S}.  We estimate the effective temperature based on an average of $V$-$J$, $V$-$H$, $V$-$K$, and $J$-$K$ colors and temperature-color relations established by \citet{2000AJ....119.1448H} and \citet{2008MNRAS.389..585C}.  A bolometric correction is computed as per \citet{2000AJ....119.1448H}, assuming solar metallicity and surface gravity $\log g = 4.5$.  Absolute magnitudes are calculated assuming a position on the main sequence \citep{2005AJ....130.1680L}.  We make three separate estimates of the mass based on the near-infrared absolute magnitudes \citep{2000A&A...364..217D,2008Ap&SS.314...51X} and average them.  Although a radius can be calculated from the Stefan-Boltzmann relation, we instead empirically infer it from the mass using the relation $R_* \approx M_*^{1.06}$ ($0.3 M_{\odot} < M_* < 0.6M_{\odot}$) based on theoretical models \citep{1998A&A...337..403B} and supported by the available data on radius measurements of single stars \citep{2009A&A...505..205D}.

We use the planet mass distribution $dn/d \log M_p = 0.39 M_p^{0.48}$ ($M$ in Earths) estimated by \citet{2010Sci...330..653H} and the period distribution of \citet{2008PASP..120..531C}: $dn/d \log $(P)$\sim$P$^{0.26}$ for $P=2-2000$~days.  These distributions are only known for solar-mass stars and may be different for M dwarfs \citep{2007ApJ...669..606R, 2009Icar..202....1M}.  The recently released {\it Kepler} catalog of planet candidates contains $\sim$1200 objects with estimated radii, but $\lesssim$3\% of them are around M dwarfs.  We adopt the orbital eccentricity distribution derived by \citet{2008ApJ...685..553S}, recognizing that the distribution is well-constrained only for giant planets and that of low-mass planets may be significantly different \citep{2009IJAsB...8..175P, 2010ApJ...719.1454M}.  Inclination angles and longitudes of periapsis are drawn from isotopic distributions.  For a mass-radius relationship we use the surprisingly simple formula $R_p = M_p^{0.5}$ (Earth units), which captures the overall trend for exoplanets of masses intermediate to those of Earth and Saturn.  This diverges from the mass-radius relationship of rocky planets derived from interior models \citep[e.g.,][]{2007ApJ...669.1279S}, possibly because of the tendency of more massive planets to have H/He envelopes.  We assume that all planets above a mass of 125 Earths are the size of Jupiter (11.2 Earth radii).

The simulated observing program is constructed as follows. Days are first selected randomly from throughout the year.  Sunset and sunrise are determined for the Mauna Kea summit, and observing begins and ends 30 minutes after and before sunset and sunrise, respectively.  The short twilight interval is justified because our targets are bright and Rayleigh scattering from the sky is low in the $z$ passband.  A target is selected at random from an updated subset of the catalog that is found at an air mass below 1.4.  We allow 2 minutes for the telescope to slew to each target and the instrument to set up, and we assume a fixed integration time of 2 minutes.  Each target that is observed is placed on hold and will not be reobserved until at least 2 hours has elapsed (the typical duration of a transit).  We do not consider the requirement that each target be observed a certain number of times each night because we ultimately sum the probability of detecting a transit over all observations, not all individual targets.

Detection requires a S/N ratio greater than 5 in a single observation.  The transit signal is calculated using the limb-darkening model of \citet{2000A&A...363.1081C}.  The total noise consists of shot noise, scintillation noise, and other (atmospheric) noise (0.4~mmag) added in quadrature.  The first two contributions are based on the performance of the SNIFS instrument and assume a SDSS~$z$ passband (see \S~\ref{sec:results} for details).  The detection probability is the product of the transit probability times the probability of detecting the transit averaged over 30,000 Monte Carlo realizations of planets.  We can impose a Doppler detection criterion by performing the average only over those systems where the radial velocity amplitude exceeds a specified value.  This simulates a scenario in which targets are first selected by Doppler observations
for transit searches.

\begin{figure}
\centering
  \includegraphics[width=\textwidth]{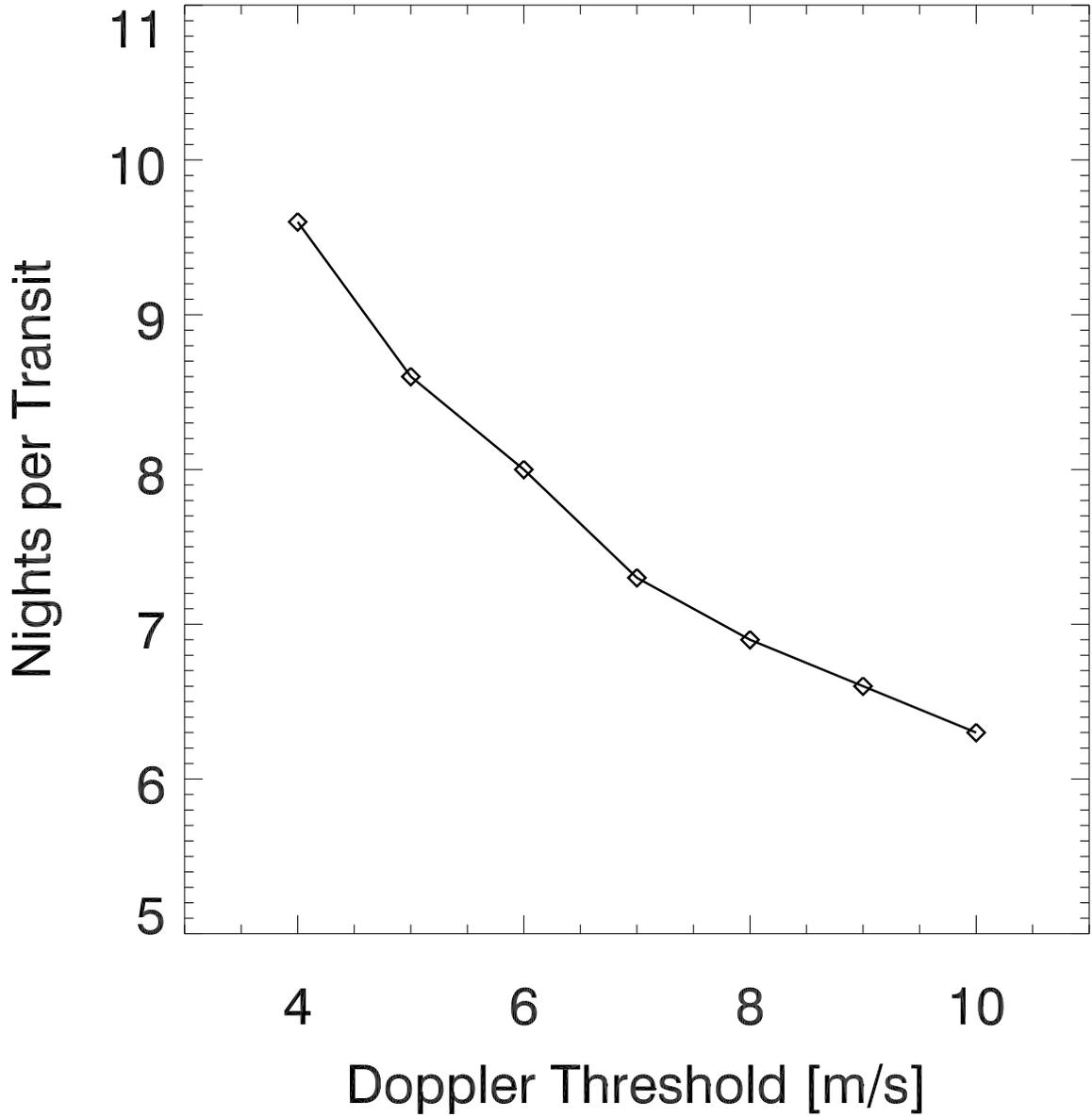}
  \caption{Average number of photometric observing nights before a transit is detected in a snapshot survey of M dwarf  stars with a Doppler signal above the specified threshold.  In making this calculation, we take into account setup, slew, twilight, and integration time, and we use precision levels attained in our SNIFS observations.  A planet population similar to those around G-type stars is assumed (see \S~\ref{sec:transitsearch} for details)}.\label{fig:plotnights}
\end{figure}

Fig.~\ref{fig:plotnights} plots the average required number of clear nights per transit discovery as a function of the Doppler threshold.  A higher threshold selects more massive planets on closer orbits; these are more likely to transit and are more readily detected if they do.  For a 6 m~s$^{-1}$ threshold, the mass distribution of detected planets peaks at around 24 $M_E$ (Neptune size).  If 56\% of nights are photometric \citep{2009PASP..121..295S}, then a 1~yr survey of 6~m~s$^{-1}$ systems would produce 25 transiting systems.  This would require a prohibitive number of Doppler-selected targets.  In the absence of Doppler selection, a completely blind survey requires a more daunting average of 90 nights per detection.

\acknowledgments
This work was supported by NSF grant AST-0908419 (to E.G.), NASA grant NNX10AI90G (to E.G.), and US Department of Energy contract DE-AC02-05CH11231 (to G.A.).  We thank John Johnson for providing the specifications of the narrow $z$ filter currently in use on the University of Hawaii 2.2~m Orthogonal Parallel Transfer Imaging Camera.  We thank Jean-Charles Cuillandre and Herb Woodruff and the rest of the Canada-France-Hawaii Telescope personnel for providing the CFHT SkyProbe data.  We thank George Ricker for his diligence and comments on the paper.  We also thank the anonymous reviewers for their helpful suggestions.

{\it Facilities:} \facility{MKO}, \facility{UH2.2m (SNIFS)}

\bibliography{/Users/amann/Dropbox/fullbiblio}

\clearpage

\begin{deluxetable}{llllll}
\tabletypesize{\scriptsize}
\tablecaption{Observing Conditions\label{tab:table1}}
\tablewidth{0pt}
\tablehead{
\colhead{UTC date} & \colhead{Seeing$^a$} & \multicolumn{3}{c}{Extinction$^b$ (mag)} & \colhead{Mean PWV$^c$ }\\
\colhead{} & \colhead{(\arcsec)} & \colhead{Median} & \colhead{std dev} & \colhead{95\% value} & \colhead{(mm)}
}
\startdata
2010 Jun 22&0.75&0.187&0.026&0.238& 1.05 \\
2010 Jun 24&0.78, 1.07& 0.198 & 0.045 & 0.240 & 1.13\\
2010 Jun 25&0.61 & 0.180 & 0.082 & 0.241 & 1.15\\
2010 Jun 27&0.99, 0.74, 0.81, 0.84& 0.225 & 0.197 & 0.942 & 2.48\\
2010 Jun 29&0.87, 0.71& 0.237 & 0.198 & 0.894 & 1.83\\
2010 Jun 30&0.68, 0.73& 0.237 & 0.202 & 0.949 & 1.43\\
2010 Jul 27&0.73& 0.213 & 0.073 & 0.319 & 1.55\\
2010 Jul 30&1.02, 1.03& 0.207 & 0.093 & 0.248 & 0.85\\
2010 Sep 15&0.96, 1.12&0.215&0.173&0.922&1.07\\
2011 Jan 15&0.83, 0.88&0.183 &0.056 &0.221 &1.74\\
\enddata
\tablenotetext{a}{From SNIFS focus images.}
\tablenotetext{b}{CFHT SkyProbe (V band) values.}
\tablenotetext{c}{Based on mean of Caltech Submillimeter Observatory 225 GHz tau measurements.}
\end{deluxetable}

\begin{deluxetable}{llrrrrrrl}
\tabletypesize{\scriptsize}
\tablecaption{Photometry Experiments\label{tab:table2}}
\tablewidth{0pt}
\tablehead{
\colhead{UTC Date} & \colhead{Experiment$^{c}$}& \colhead{Integrations} & \colhead{Cadence} & \colhead{Aper. rad.} & \multicolumn{2}{c}{Signal$^a$} & rms$^{b}$ \\
\colhead{} & & \colhead{(sec)} & \colhead{(min)} & \colhead{(pixels)} & \colhead{Target} & \colhead{Reference} & $\times10^{-4}$ &
}
\startdata
2010 Jul 27 & Snapshot (8-star) & 9 $\times$ 19 & 17 & 80 & 15.8 & ... &  5.4 (5.8) & \\
2010 Jul 27 & Snapshot (8-star) & 9 $\times$ 31 & 17 & 75 & 11.7 & ... & 6.0 (7.0) & \\
2010 Jul 27 & Snapshot (8-star) & 9 $\times$ 22 & 17 & 90 & 15.7 & ... & 8.9 (5.2) & \\
2010 Jul 27 & Snapshot (8-star) & 9 $\times$ 30 & 17 & 75 & 14.2 & ... & 7.3 (8.0) & \\
2010 Jul 27 & Snapshot (8-star) & 9 $\times$ 13 & 17 & 80 & 11.8 & ... & 8.3 (9.0) & \\
2010 Jul 27 & Snapshot (8-star) & 9 $\times$ 23 & 17 & 85 & 16.3 & ... & 7.8 (6.9) & \\
2010 Jul 27 & Snapshot (8-star) & 9 $\times$ 13 & 17 & 75 & 13.6 & ... & 6.5 (7.5) & \\
2010 Jul 27 & Snapshot (8-star) & 9 $\times$ 18 & 17 & 75 & 14.0 & ... & 8.4 (8.0) & \\
2010 Jul 31 & Common field & 15 $\times$ 88 & 2.4 & 65, 80 & 8.4 & 13.6 & 9.7 (---) &\\
2010 Jul 27 & Common field & 13 $\times$ 22 & 5.5 & 60, 85 & 13.1 & 12.5 & 8.4 (---) & \\
2011 Jan 15 & $r$ Filter test & 20 $\times$ 50 & 9.6 & 85 & 17.1 & 16.8 & 9.2 (9.5) & \\
2011 Jan 15 & $z$ Filter test & 20 $\times$ 50 & 9.6 & 80 & 18.0 & 17.6 & 8.5 (7.5) & \\
2011 Jan 15 & Exposure time tests & 22 $\times$ 25 & 18.1 & 65 & 8.0 & 7.1 & 11.5 (11.7) & \\
2011 Jan 15 & Exposure time tests & 22 $\times$ 50 & 18.1 & 75 & 16.1 & 14.0 & 9.8 (10.1) & \\
2011 Jan 15 & Exposure time tests & 22 $\times$ 100 & 18.1 & 90 & 31.9 & 27.8 & 8.0 (11.5) & \\
\enddata
\tablenotetext{a}{$10^6$~e$^{-}$ per integration.}
\tablenotetext{b}{Value in parentheses is theoretical precision (see \S~\ref{sec:noisemodel} and Table \ref{tab:noisetable}).}
\tablenotetext{c}{With the exception of the January 15 filter tests, all data listed here were taken with the $z$ filter.}
\end{deluxetable}

\begin{deluxetable}{llrrrrlll}
\tabletypesize{\scriptsize}
\tablecaption{Estimated Error/Noise Budget\label{tab:noisetable}}
\tablewidth{0pt}
\tablehead{
\colhead{Noise Source$^a$}  & \colhead{Median} & \colhead{Mean}  \\
\colhead{} &\colhead{$10^{-4}$} & \colhead{$10^{-4}$} } 
\startdata
Poisson&4.8&4.9\\
Scintillation&3.4&3.3\\
Atmospheric transparency fluctuations&2.8&2.9\\
First-order extinction&1.0&0.91\\
Sky&0.99&1.3\\
Read and electronic pickup&0.86&0.93\\
Motion on chip&0.85&1.0\\
Second-order extinction &0.45&0.56\\
Shape changes&0.27&0.34\\
Chip/Amplifier nonlinearities&0.020&0.021\\
\enddata
\tablenotetext{a}{Noise for each star is based on our model applied to each observation (theoretical and actual precision shown in Fig.~\ref{fig:theoretical}).  Noise is calculated per light curve (accounting for target and all comparison stars).}
\end{deluxetable}

\clearpage

\end{document}